\documentclass[seceq,preprint]{ptptex}

\usepackage{epsfig, color,graphicx}
\usepackage{amsmath}
\usepackage{amssymb}
\newcommand{\nn}{\nonumber \\}
\newcommand{\bea}{\begin{eqnarray}}
\newcommand{\ena}{\end{eqnarray}}

\newcommand{\hs}[1]{\hspace{#1 mm}}
\renewcommand{\a}{\alpha}

\renewcommand{\c}{\gamma}
\renewcommand{\d}{\delta}

\newcommand{\s}{\sigma}

\newcommand{\la}{\lambda}

\newcommand{\p}[1]{(\ref{#1})}
\newcommand{\tm}{\tilde{m}}
\newcommand{\tr}{\tilde{r}}

\newcommand{\squaret}{\kern1pt\vbox{\hrule height 0.9pt\hbox{\vrule width
0.9pt\hskip 2pt\vbox{\vskip 5.5pt}\hskip 3pt\vrule width 0.3pt}\hrule height
0.3pt}\kern1pt}
\begin{document}

\preprintnumber[3cm]{
KU-TP 027}
%%
%% title
%%
\title{\large  Black Holes in the Dilatonic Einstein-Gauss-Bonnet Theory
in Various Dimensions III \\
-- Asymptotically AdS Black Holes with $k=\pm 1$ --
}
%%
%% author
%%
\author{
Nobuyoshi {\sc Ohta}$^{a,}$\footnote{e-mail address: ohtan at phys.kindai.ac.jp}
and Takashi {\sc Torii}$^{b,}$\footnote{e-mail address: torii at ge.oit.ac.jp}
}

\inst{
$^a$Department of Physics, Kinki University, Higashi-Osaka,
Osaka 577-8502, Japan\\
$^b$Department of General Education, Osaka Institute of Technology,
Asahi-ku, Osaka 535-8585, Japan
}
%%
%% abstract
%%
\abst{
We study black hole solutions in the Einstein-Gauss-Bonnet gravity with the dilaton
and a negative ``cosmological constant''.
We derive the field equations for the static spherically symmetric ($k=1$)
and hyperbolically symmetric ($k=-1$) spacetime in general $D$ dimensions.
The system has some scaling symmetries which are used in our analysis of the solutions.
We find exact solutions, i.e., regular AdS solution for $k=1$ and a massless
black hole solution for $k=-1$. Nontrivial asymptotically AdS solutions are
obtained numerically in $D=4$ -- 6 and 10 dimensional spacetimes.
For spherically symmetric solutions, there is the minimum horizon radius
below which no solution exists in $D=4$ -- 6. However in $D=10$,
there is not such lower bound but the solution continues to exist
to zero horizon size.
For hyperbolically symmetric solution, there is the minimum horizon radius
in all dimensions.
Our solution can be used for investigations of the boundary theory
through AdS/CFT correspondence.

%-----------------------------------------

%\begin{center}
%\begin{tabular}{c|cccc}
%\hline
% & $~~k=1~~$ & $~~k=0~~$ & $k=-1$ (GR) & $k=-1$ (GB) \\
%\hline
%$\Lambda=0$ & $\bigcirc$ & $\times$  & &  \\
%$\Lambda=1$ &  &   & &  \\
%$\Lambda=-1$ & $\spadesuit$ & $\bigcirc$  &$\spadesuit$ & $\spadesuit$ \\
%\hline
%\end{tabular}
%\end{center}
}

\maketitle

%%%%%%%%%%%%%%%%%%%%%%%%%%%%%%%%%%%%
%%%%%%%%%%%%%%%%%%%%%%%%%%%%%%%%%%%%
\section{Introduction}
%%%%%%%%%%%%%%%%%%%%%%%%%%%%%%%%%%%%
%%%%%%%%%%%%%%%%%%%%%%%%%%%%%%%%%%%%

This is the third of the series of our papers about the black hole solutions
in dilatonic Einstein-Gauss-Bonnet theory in higher dimensions.~\cite{GOT1,GOT2}

Many works have been done on black hole solutions in dilatonic gravity,
and various properties have been studied since the works in Refs.~\citen{GM} and
\citen{GHS}.
It was found that the dilaton modifies the properties of the black hole solution.
On the other hand, it is known that there are higher-order corrections from
string theories.~\cite{GS} It is then natural to ask how these corrections may modify
the results. Several works have studied the effects of higher order
terms,~\cite{KMRTW,AP,TYM,CGO1,CGO2}
but most of the work done so far considers theories without dilaton,~\cite{BD,GG,Cai}
which is one of the most important ingredients in the string effective theories.
Hence it is important to study black hole solutions and their properties
in the theory with both the higher order corrections and dilaton.
The simplest higher order correction is the Gauss-Bonnet (GB) term,
which may appear in heterotic string theories.

In the first paper in this series~\cite{GOT1}, we focused on asymptotically
flat solutions and studied system with the GB correction term and dilaton without
the cosmological term in various dimensions from 4 to 10.
They are spherically symmetric with the $(D-2)$-dimensional
hypersurface of curvature signature $k=1$.
We have then turned to planer symmetric ($k=0$) topological black holes,
but have found that no solution exists without the cosmological term.
In the string perspective, it is more interesting to examine asymptotically
anti-de Sitter (AdS) black hole solutions with possible application to AdS/CFT
correspondence in mind.
Discussions of the origin of such cosmological terms are given in Refs.~\citen{AGMV,POL}.
So in the second paper in this series,~\cite{GOT2} we have presented the results
on black hole solutions with a negative cosmological term with $k=0$.
In fact, shear viscosity has been computed using the result in our previous
paper.~\cite{CNOS}

In this paper, we extend our previous work to solutions with $k=\pm 1$ and a negative
cosmological term.
Black hole solutions in dilatonic Einstein-Maxwell theories with Liouville-type
potential but without GB term are studied in Refs.~\citen{PW,CHM}.
Exact solutions and their properties are discussed in Ref.~\citen{Ch}
in dilatonic Einstein theory with Liouville potential.
Cosmological solutions are also considered in Ref.~\citen{BGO}.
The presence of a Liouville potential already changes completely
the difficulty of the system.
In fact it is no longer an integrable system even in the absence of a GB term
and our system is far much difficult one.
Nevertheless, we find some exact solutions in our complicated system of Einstein-GB
system with a dilaton. These are the regular AdS solution for $k=1$ and
a massless black hole solution for $k=-1$.
We also obtain nontrivial asymptotically AdS solutions numerically in $D=4$ -- 6 and
10 dimensional spacetimes and discuss their properties.
We expect that these solutions should be useful for studying properties
of dual filed theories.

This paper is organized as follows. In \S~2, we first present the action
and give basic equations to solve.
Symmetry properties of the theory are also discussed in order to apply them
in our following analysis.
In \S~3, boundary conditions at the horizon and relevant asymptotic behaviors are
examined.
In \S~4 and \S~5, we first give the exact solutions of the regular AdS solution
for $k=1$ and a massless black hole solution for $k=-1$, and then present our
numerical solutions in $D=4$ -- 6 and 10 for $k=+1$ and $k=-1$, respectively.
These solutions are presented for a particular choice of the parameters in our theory,
but we expect that the qualitative behaviors do not change for other choices
though the range of the horizon radii for the existence of the black hole solutions
seems to change depending on the strength of the dilaton coupling $\c$.
Some discussions on this problem as well as on the difference between string frame
and Einstein frame is given in Ref.~\citen{MOS} for asymptotically flat solutions
without cosmological term.
\S~6 is devoted to conclusions and discussions.

%%%%%%%%%%%%%%%%%%%%%%%%%%%%%%%%%%%%
%%%%%%%%%%%%%%%%%%%%%%%%%%%%%%%%%%%%
\section{Dilatonic Einstein-Gauss-Bonnet theory}
%%%%%%%%%%%%%%%%%%%%%%%%%%%%%%%%%%%%
%%%%%%%%%%%%%%%%%%%%%%%%%%%%%%%%%%%%

%%%%%%%%%%%%%%%%%%%%%%%%%%%%%%%%%%%%
\subsection{Action and basic equations}
%%%%%%%%%%%%%%%%%%%%%%%%%%%%%%%%%%%%

We consider the following low-energy effective action for a
heterotic string
\bea %----------------
S=\frac{1}{2\kappa_D^2}\int d^Dx \sqrt{-g} \left[R - \frac12
 (\partial_\mu \phi)^2
 + \a_2 e^{-\c \phi} R^2_{\rm GB} -\Lambda e^{\la \phi} \right],
\label{act}
\ena %----------------
where $\kappa_D^2$ is a $D$-dimensional gravitational constant,
$\phi$ is a dilaton field, $\alpha_2=\a'/8$ is a numerical
coefficient given in terms of the Regge slope parameter,
and
$R^2_{\rm GB} = R_{\mu\nu\rho\sigma} R^{\mu\nu\rho\sigma}
- 4 R_{\mu\nu} R^{\mu\nu} + R^2$ is the GB correction.
We leave the coupling constant of dilaton $\gamma$ arbitrary as much as possible,
while the ten-dimensional critical string theory predicts $\c=1/2$.
Since the negative cosmological constant $\Lambda=-(D-1)(D-2)/\ell^2$
couples to the dilaton with the coupling constant $\la$, this term can be
regarded as the potential of the dilaton.
If this is the only potential, there is no stationary point and the dilaton
cannot have a stable asymptotic value. However, for asymptotically
AdS solutions, the Gauss-Bonnet term produces an additional potential
in the asymptotic region, and the dilaton can take finite constant value at infinity.
There may be several possible sources of ``cosmological terms'' with different
dilaton couplings, so we leave $\lambda$ arbitrary and specify it
in the numerical analysis.

Varying the action~\p{act} with respect to $g_{\mu\nu}$, we obtain
the gravitational equation:
\begin{eqnarray}
\label{GB-eq}
&& G_{\mu\nu}
-\frac12\biggl[\nabla_{\mu}\phi\nabla_{\nu} \phi -\frac12 g_{\mu\nu}(\nabla\phi)^2\biggr]
+\frac12 g_{\mu\nu}\Lambda e^{\la\phi}
\nonumber \\
&& \hspace{10mm}
+\alpha_2 e^{-\gamma\phi}\Bigl[H_{\mu\nu}
+4(\gamma^2\nabla^{\rho}\phi\nabla^{\sigma}\phi
-\gamma\nabla^{\rho}\nabla^{\sigma}\phi)P_{\mu\rho\nu\sigma}\Bigr]
=0,
\end{eqnarray}
where
\begin{eqnarray}
&&G_{\mu\nu}\equiv R_{\mu\nu}-{1\over 2}g_{\mu\nu}R,
\\
&&H_{\mu\nu}\equiv 2\bigl(RR_{\mu\nu}-2R_{\mu \rho}R^{\rho}_{~\nu}
-2R^{\rho\sigma}R_{\mu\rho\nu\sigma}
+R_{\mu}^{~\rho\sigma\lambda}R_{\nu\rho\sigma\lambda}\bigr)
-{1\over 2}g_{\mu\nu}R^2_{\rm GB},
\\
&& P_{\mu\nu\rho\sigma}\equiv
R_{\mu\nu\rho\sigma}+2g_{\mu[\sigma}R_{\rho]\nu}
+2g_{\nu[\rho}R_{\sigma]\mu} +Rg_{\mu[\rho}g_{\sigma]\nu}.
\label{EGB:eq}
\end{eqnarray}
$P_{\mu\nu\rho\sigma}$ is the divergence free part of the Riemann tensor,
i.e.
$\nabla_\mu P^{\mu}_{~\nu\rho\sigma}=0$.
The equation of the dilaton field is
\begin{eqnarray}
\label{dil-eq}
\squaret \phi -\alpha_2 \gamma e^{-\gamma\phi}  R^2_{\rm GB}
-\la \Lambda e^{\la\phi}=0,
\end{eqnarray}
where $\squaret$ is the $D$-dimensional d'Alembertian.

We parametrize the metric as
\bea
ds_D^2 = - B e^{-2\d} dt^2 + B^{-1} dr^2 + r^2 h_{ij}dx^i dx^j,
\ena
where $h_{ij}dx^i dx^j$ represents the line element of a
$(D-2)$-dimensional hypersurface with constant curvature
$(D-2)(D-3)k$ and volume $\Sigma_k$ for $k=\pm 1,0$.

The metric function $B=B(r)$ and the lapse function $\d=\d(r)$ depend only on the
radial coordinate $r$. The field equations are~\cite{GOT2}
\bea
&& \bigl[(k-B)\tr^{D-3}\bigr]' \frac{D-2}{\tr^{D-4}}h -\frac12 B \tr^2 {\phi'}^2
 - (D-1)_4\,e^{-\c\phi}\frac{(k-B)^2}{\tr^2} \nn
&& \hs{10} + 4(D-2)_3\, \c e^{-\c\phi}B(k-B)(\phi''-\c {\phi'}^2) \nn
&& \hs{10} + 2(D-2)_3\,\c e^{-\c\phi}\phi'\frac{(k-B)[(D-3)k-(D-1)B]}{\tr}
-\tr^2 \tilde \Lambda e^{\la \phi}= 0\,,
\label{fe1} \\
&& \delta'(D-2)\tr h + \frac12 \tr^2 {\phi'}^2
 -2(D-2)_3\, \c e^{-\c\phi}(k-B)(\phi''-\c {\phi'}^2) =0 \,,
 \label{fe2} \\
&&
(e^{-\d} \tr^{D-2} B \phi')' = \c (D-2)_3 e^{-\c\phi-\d} \tr^{D-4}
\Big[ (D-4)_5 \frac{(k-B)^2}{\tr^2} + 2(B'-2\d' B)B' \nn
&& \hs{10} -4(k-B)BU(r)
-4\frac{D-4}{\tr}(B'-\d'B)(k-B) \Big] + e^{-\d} \tr^{D-2} \la \tilde\Lambda e^{\la\phi},
\label{fe3}
\ena
where $\tr \equiv r/\sqrt{\a_2}$,
$\tilde \Lambda = \a_2 \Lambda$ are
the dimensionless variables, and the primes in the field equations
denote the derivatives with respect to $\tr$. Other functions are defined as
\bea
(D-m)_n &\equiv& (D-m)(D-m-1)(D-m-2)\cdots(D-n), \nn
\label{h-def}
h &\equiv& 1+2(D-3) e^{-\c\phi} \Big[ (D-4) \frac{k-B}{\tr^2}
 + \c \phi'\frac{3B-k}{\tr}\Big], \\
\label{tilh-def}
\tilde h &\equiv& 1+2(D-3) e^{-\c\phi} \Big[(D-4)\frac{k-B}{\tr^2}
+\c\phi'\frac{2B}{\tr} \Big], \\
U(r) &\equiv& (2 \tilde h)^{-1} \Bigg[ (D-3)_4 \frac{k-B}{\tr^2 B}
-2\frac{D-3}{\tr}\Big(\frac{B'}{B}-\d'\Big) -\frac12 \phi'^2 \nn
&& + (D-3)e^{-\c\phi} \Bigg\{ (D-4)_6 \frac{(k-B)^2}{\tr^4 B}
- 4 (D-4)_5 \frac{k-B}{\tr^3}\Big(\frac{B'}{B}-\d'-\c\phi'\Big) \nn
&& -4(D-4)\c \frac{k-B}{\tr^2}\Big( \c \phi'^2 +\frac{D-2}{\tr}\phi'-\Phi \Big)
\nn
&&
+8 \frac{\c\phi'}{\tr} \biggl[\Big(\frac{B'}{2}-\d' B\Big)\Big(\c\phi'-\d'
+\frac{2}{\tr} \Big) -\frac{D-4}{2\tr}B' \biggr]
\nn
&&
+ 4(D-4)\Big(\frac{B'}{2B}-\d' \Big)
\frac{B'}{\tr^2}-\frac{4\c}{\tr}\Phi (B'-2\d'B)\Bigg\}
-\frac{1}{B} \tilde \Lambda e^{\la \phi}\Biggr],\\
\Phi &\equiv& \phi'' +\Big(\frac{B'}{B}-\d' +\frac{D-2}{\tr}\Big) \phi'.
\label{dil}
\ena

%%%%%%%%%%%%%%%%%%%%%%%%%%%%%%%%%%%%
\subsection{Symmetry and scaling}
%%%%%%%%%%%%%%%%%%%%%%%%%%%%%%%%%%%%

Our field equations has several scaling symmetries.
Firstly the field equations are invariant under the transformation:
\bea
\gamma \to -\gamma, ~~
\lambda \to -\lambda, ~~
\phi \to -\phi\, .
\label{sym1}
\ena
By this symmetry, we can restrict the parameter range of $\c$ to $\c\geq 0$.
The second one is the  shift symmetry:
\bea
\phi \to \phi-\phi_{\ast}, ~~
\tilde\Lambda \to e^{(\la-\c)\phi_{\ast}} \tilde\Lambda, ~~
r \to e^{-\gamma\phi_{\ast}/2}r.
\label{sym2}
\ena
where $\phi_{\ast}$ is an arbitrary constant.
This may
be used to generate solutions for different cosmological constants,
given a solution for some cosmological constant and $\tr_H$.
Details will be discussed in \S~4.
The final one is the shift symmetry under
\bea
\delta \to \delta - \delta_{\ast}, ~~
t \to  e^{-\delta_{\ast}}t,
\label{sym3}
\ena
with an arbitrary constant $\delta_{\ast}$, which may be used to shift
the asymptotic value of $\delta$ to zero.

The model \p{act} has several parameters of the theory such as
$D$, $\alpha_2$, $\Lambda$, $\gamma$, and $\lambda$. The black hole solutions
have also physical parameters such as the horizon radius $\tr_H$ and
the value of $\delta$ at infinity.
However owing to the above symmetries (including the scaling by $\alpha_2$),
we can reduce the number of the parameters and are left only with
$D$, $\gamma\geq 0$, $\lambda$, and $\tr_H$.

%%%%%%%%%%%%%%%%%%%%%%%%%%%%%%%%%%%%
%%%%%%%%%%%%%%%%%%%%%%%%%%%%%%%%%%%%
\section{Boundary conditions and asymptotic behavior}
%%%%%%%%%%%%%%%%%%%%%%%%%%%%%%%%%%%%
%%%%%%%%%%%%%%%%%%%%%%%%%%%%%%%%%%%%

In the numerical search for the appropriate black hole solutions, we should
impose the conditions that they have the regular event horizons and
outer spacetimes should be singularity free.
As for the asymptotic structure, we assume AdS like
spacetime. Let us now examine them in detail.

%%%%%%%%%%%%%%%%%%%%%%%%%%%%%%%%%%%%
\subsection{Regular horizon}
%%%%%%%%%%%%%%%%%%%%%%%%%%%%%%%%%%%%

We impose the following boundary conditions for the metric functions:
\begin{enumerate}
\item
The existence of a regular horizon $\tr_H$:
\bea
\label{hor}
B(\tr_H)=0, ~~
|\d_H| < \infty, ~~
|\phi_H|< \infty\, .
\ena
\item
The nonexistence of singularities outside the event horizon ($\tr > \tr_H$):
\bea
B(\tr)>0, ~~
|\d| < \infty, ~~
|\phi|< \infty\, .
\ena
\end{enumerate}
Here and in what follows, the values of various quantities at the horizon are
denoted with subscript $H$.
At the horizon, it follows from Eqs.~\p{fe1}, \p{h-def}, and \p{tilh-def} that
\bea
&& B_H=0, \nn
&&
h_H = 1+2(D-3)k\frac{e^{-\c\phi_H}}{\tr_H^2}[D-4 -\c\phi_H' \tr_H],\nn
&&
\tilde h_H= 1+2(D-3)_4k \frac{e^{-\c\phi_H}}{\tr_H^2},\nn
&& B_{H}' h_H = \frac{(D-3)k}{\tr_H} +\frac{(D-3)_5k^2}{\tr_H^2} e^{-\c\phi_H}
- \tr_H \frac{\tilde\Lambda e^{\la\phi_H}}{D-2}.
\label{bhor}
\ena
Using these in Eq.~\p{fe3} gives
\bea
&&
2kC\c\Big[-2  L \tr_H^2  \Big\{ 1 + kC \big[D-4 -(D-2) \c \la\big]
+  (D-2)k^2C^2 \c \big[(D-6) \c -(D-4) \la \big] \Big\} \nn
&& \hs{10}
+  (D-2)k \Big\{ 2(D-3) + (D-4)(3 D-11) kC \nn
&& \hs{15}
+ (D-4) k^2C^2 \big[ (D-2)(3D-11) \c^2+ (D-4)_5 \big] +2(D-2)_5 k^3C^3 \c^2 \Big\}\Big]
\tr_H^2 \phi_H'^2 \nn
%--
&&
+ 2\Big[4k^2C^2 \c ^2L^2 \tr_H^4
+ 2 L \tr_H^2 \Big\{ 1+ 2kC\big[D-4-(D-2)\c\la\big]
\nn
&& \hs{20}
-k^2C^2\big[ 2(D-2)(2D-5)\c^2
+4(D-2)(D-4)\c\la-(D-4)^2 \big]
\nn
&& \hs{20}
-2(D-2) (D-4) k^3C^3 \c\big[(D-2)\c+(D-4)\la \big] \Big\} \nn
&&  \hs{10}
+(D-2) k\Big\{  (D-1)_2 (D-4) k^2C^2 \c^2\big[ 2+2kC-(D-4)_5 k^2C^2 \big] \nn
&& \hs{30}
- \big[1+(D-4)kC \big]^2 \big[2(D-3)+(D-4)_5 kC \big]\Big\} \Big]\tr_H\phi_H' \nn
%--
&&
+ 4 C\c L^2 \tr_H^4 + 4 (D-2) L \tr_H^2 \Big\{ \la \big[ 1+(D-4)kC \big]^3 \nn
&& \hs{30}
+ kC\c \big[ D(D-4)^2 k^2C^2 + (D-4)(D+1) kC -(D-2) \big] \Big\} \nn
&& \hs{5}
+ (D-1)_2 (D-2) k^2C\c \Big\{ 2(D-2) -4 (D-4) kC - (D-4)^2 (D+1) k^2C^2 \Big\}=0,
\label{pder}
\ena
%---"ñ•\Ž¦---------------------------------------------------------------------------
\if0
{\color{blue}
\bea
&&
\Big[-4 C L \tr_H^2 \c \Big\{ 1 + C \Big(D-4 -(D-2) \c \la\Big)
+ C^2 (D-2) \c \Big((D-6) \c -(D-4) \la \Big) \Big\} \nn
&& \hs{10}
+ 2 (D-2) C \c \Big\{ 2(D-3) + (D-4)(3 D-11) C \nn
&& \hs{20}
+ (D-4) C^2 \Big( (D-2)(3D-11) \c^2+ (D-4)_5 \Big) +2(D-2)_5 C^3 \c^2 \Big\}\Big]
\tr_H^2 \phi_H'^2 \nn
&&
+ \Big[8 C^2 L^2 \tr_H^4 \c ^2 + 4 L \tr_H^2 \Big\{ 1+ 2\Big(D-4-(D-2)\c\la\Big) C
- \Big( 2(D-2)(2D-5)\c^2 \nn
&&
+4(D-2)(D-4)\c\la-(D-4)^2 \Big) C^2
-2(D-2) (D-4) \Big((D-2)\c+(D-4)\la \Big) C^3 \c \Big\} \nn
&&
+2(D-2) \Big\{ C^2 \c^2 (D-1)_2 (D-4) \Big( 2+2C-(D-4)_5 C^2 \Big) \nn
&& \hs{30}
- \Big(1+(D-4)C \Big)^2 \Big(2(D-3)+(D-4)_5 C \Big)\Big\} \Big]\tr_H\phi_H' \nn
&&
+ 4 C L^2 \tr_H^4 \c + 4 (D-2) L \tr_H^2 \Big[ \la \Big( 1+(D-4)C \Big)^3 \nn
&& \hs{30}
+\c C \Big( D(D-4)^2 C^2 + (D-4)(D+1) C -(D-2) \Big) \Big] \nn
&&
+ (D-1)_2 (D-2) C\c \Big[ 2(D-2) -4 (D-4) C - (D-4)^2 (D+1) C^2 \Big]=0,
%\label{pder}
\ena
}
\fi
%-----------------------------------------------------------------------------------------
where we have defined
\bea
C=\frac{2(D-3)e^{-\c\phi_H}}{\tr_H^2},~~
L = e^{\la\phi_H} \tilde\Lambda.
\label{c}
\ena
Eq.~\p{pder} is a quadratic equation to determine $\phi'_H$ and there can be solutions
only if we have real solutions in this equation. We will see that this gives
strong constraint on our solutions.

%%%%%%%%%%%%%%%%%%%%%%%%%%%%%%%%%%%%
\subsection{Asymptotic behavior at infinity}
%%%%%%%%%%%%%%%%%%%%%%%%%%%%%%%%%%%%

At infinity we impose the condition that the leading term of the metric function
$B$ comes from AdS radius, i.e.,
\begin{enumerate}
\item[3.]
``AdS asymptotic behavior" ($\tr \to \infty$):
\bea
\label{as}
B \sim \tilde{b}_2 \tr^2 +k - \frac{2\tilde M}{\tr^{\mu}}, ~~~
\d(r) \sim \d_0 + \frac{\d_1}{\tr^{\s}}, ~~
\phi \sim \phi_0 + \frac{\phi_1}{\tr^{\nu}} \,,
\ena
with finite constants $\tilde{b}_2>0$, $\tilde M$, $\d_0$, $\d_1$,  $\phi_0$, $\phi_1$
and positive constant $\mu$, $\s$, $\nu$.
\end{enumerate}
The coefficient of the first term $\tilde{b}_2$ is related to the AdS radius
as $\tilde{b}_2 = \tilde\ell_{\rm AdS}^{-2}$. However, this condition is not sufficient
for the spacetime to be AdS asymptotically. Strictly
speaking, asymptotically AdS spacetime is left invariant under
$SO(D-1,2)$.~\cite{Henneaux} Whether the solution satisfies the AdS-invariant
boundary condition or not depends on the value of the power indices
$\mu$, $\s$, $\nu$.

%-----------------------------------

%\subsection{Asymptotic expansion}

Substituting Eqs.~(\ref{as}) into the field equations (\ref{fe1}) and
(\ref{fe3}), one finds the leading terms ($\tr^2$ and constant terms
in each equation) give the conditions
\bea
&&
\label{leading1}
(D)_3\gamma\; e^{-\gamma\phi_0} \tilde{b}_2^{~2}
+ \lambda \tilde{\Lambda}e^{\lambda\phi_0} = 0 \,, \\
&&
\label{leading2}
(D-1)_4\;e^{-\gamma\phi_0} \tilde{b}_2^{~2}-(D-1)_2 \tilde{b}_2
-\tilde{\Lambda}e^{\lambda\phi_0} = 0 \,,
\ena
which determine $\tilde b_2$ and $\phi_0$,
while $\d_0$ can be arbitrary because only its derivative appears in the field
equations. We can restrict the parameter to $\lambda>0$ by Eq.~\p{leading1}
and assume $\lambda\ne \c$.
Then, Eqs.~(\ref{leading1}) and (\ref{leading2}) give
\bea
&&
\label{b2inf}
\tilde{b}_2^{~2}
=\frac{-\lambda\tilde{\Lambda}}{(D)_3 \gamma}
\biggl[\frac{D(D-3)}{(D-1)_2}(-\tilde{\Lambda})\frac{\gamma}{\lambda}
\biggl(1+\frac{(D-4)\lambda}{D\gamma}\biggr)^2
\biggr]^{\frac{\gamma+\lambda}{\gamma-\lambda}},
\\
&&
\label{phiinf2}
e^{\phi_0}
=\biggl[\frac{D(D-3)}{(D-1)_2}(-\tilde{\Lambda})\frac{\gamma}{\lambda}
\biggl(1+\frac{(D-4)\lambda}{D\gamma}\biggr)^2
\biggr]^{\frac{1}{\gamma-\lambda}}.
\ena

The candidates of the next leading terms are almost the same as
those in $k=0$ case except for some additional terms which are
proportional to $k$. However, these additional terms turn out to be further
subleading ones by detailed analysis. This means that the asymptotic behaviors are
the same as in the $k=0$ case, which is discussed in our previous
paper.~\cite{GOT2} Hence we only summarize the results here without details.

\if0 %-------------------------------------------------
for Eqs.~\p{fe1}-\p{fe3} are
respectively given by
\bea
&&
2(D-2)\left[\mu-(D-3)\right]\left[1-2(D-3)_4 \tilde{b}_2e^{-\c\phi_0}\right]\tilde{M}
\tr^{-\mu} \nn
&& \hs{10}
+ 4(D-2)_3 \{(D-3)-\nu \} \tilde b_2 e^{-\c\phi_0}\c \nu\phi_1 \tr^{-\nu} \nn
&& \hs{10} - \Bigl[(D-2)_3\c \tilde{b}_2^2e^{-\c\phi_0}
\{4\nu^2-4(D-2)\nu+(D-1)(D-4)\} + \la \tilde{\Lambda} e^{\la\phi_0}
\Bigr]\phi_1 \tr^{2-\nu} ,
\label{next1} \\
%
&& \label{next2} \rule[0mm]{0mm}{7mm}
\Bigl[2(D-3)_4 \tilde{b}_2 e^{-\c\phi_0}-1\Bigr]\sigma \d_1 \tr^{-\sigma}
 + 2(D-3) \tilde{b}_2 e^{-\c\phi_0}\gamma\nu(1+\nu)\phi_1 \tr^{-\nu} ,\\
%
&& \rule[0mm]{0mm}{7mm}
4(D-2)_3\tilde{b}_2^2e^{-\c\phi_0}\c \sigma (\sigma-D)\d_1 \tr^{-\sigma}
 + \Bigl[D_3\tilde{b}_2^2e^{-\c\phi_0}\c^2 - \la^2 \tilde{\Lambda} e^{\la\phi_0}
 - \tilde{b}_2(D-1)\nu + \tilde{b}_2 \nu^2 \Bigr]\phi_1 \tr^{-\nu} \nonumber \\
&& \hs{10} - 4(D-2)_3\tilde{b}_2e^{-\c\phi_0}[\mu-(D-2)][\mu-(D-3)]\c \tilde{M}
\tr^{-\mu-2} \nn
&& \hs{15}
+4(D-2)_3 \tilde b_2 e^{-\c\phi_0} \c \d_1 \s^2 \tr^{-\s-2},
\label{next3}
\ena
which should vanish. These are again similar to the results in our previous
paper,~\cite{GOT2} but there are some additional terms.
These additional terms turn out to be further subleading ones,
and we get basically the same asymptotic behaviors as the $k=0$ case.
\fi %-------------------------------------------------

We consider the case with $(D-4)\lambda-D\gamma\ne 0$.
There are two different classes which give consistent expansions.
One is  $\mu=D-3$ and $\nu, \s >D-1$, and we rename the coefficient $\tilde M$
as $\tilde M_0$.

The other class is $\mu=\nu-2=\sigma-2$ and
\bea
\nu =\nu_{\pm} \equiv \frac{D-1}2 \left[1 \pm \sqrt{1 - \frac{4(D)_2\la\c(\la-\c)
\bigl[(D-4)\la+D\c\bigr]}{(D-1)^2\bigl[(D-4)^2\la^2-D^2\c^2-8(D-1)_2\la^2\c^2\bigr]}}
\;\right].
\label{nu}
\ena

The asymptotic forms of the fields are then
\bea
&&
\phi \sim \phi_0%+ \frac{\phi_{-}}{\tr^{\nu_{-}}}
+ \frac{\phi_{+}}{\tr^{\nu_{+}}}
+ \cdots\,,\nn
&&
\label{behavior}
B \sim \tilde{b}_2 \tilde{r}^2 +k
- \frac{2 \tilde{M}_{+}}{\tr^{\nu_{+}-2}} - \frac{2\tilde{M}_0}{\tr^{D-3}} + \cdots\,,\\
&&
\delta \sim \frac{\delta_{+}}{\tr^{\nu_{+}}}+ \cdots \, .\nonumber
\ena
There will be in general $\phi_{-}/\tr^{\nu_{-}}$ term in the asymptotic
behavior of $\phi$, but we tune the boundary condition of $\phi_H$ to kill this term.
Note that while $B$ has the term $\tr^{-\nu_{+}+2}$,
the $g_{tt}$ component of the metric behaves as
\begin{eqnarray}
-g_{tt}=Be^{-2\delta} \sim \tilde{b}_2 \tilde{r}^2 +k
- \frac{2\tilde{M}_0}{\tr^{D-3}} + \cdots.
\end{eqnarray}
This value of $\tilde{M}_0$ is the gravitational mass of the black holes.
Thus it is convenient to define the mass function $\tm_g(\tr)$ by
\bea
-g_{tt} = \tilde b_2 \tr^2 +k - \frac{2\tm_g(\tr)}{\tr^{D-3}}.
\label{mass}
\ena
We will present our results in terms of this function.

%%%%%%%%%%%%%%%%%%%%%%%%%%%%%%%%%%%%
\subsection{Allowed parameter regions}
%%%%%%%%%%%%%%%%%%%%%%%%%%%%%%%%%%%%

As in our previous paper,~\cite{GOT2} imposing the conditions
\begin{equation}
\tilde{m}_{BF}^2\leq \tilde{m}^2,~~
\tilde{m}^2<0,
\label{BFbound}
\end{equation}
where~\cite{BF,HM}
\bea
\tilde{m}_{BF}^2&&=-\frac{(D-1)^2}{4\tilde{\ell}_{AdS}^2}
=-\frac{(D-1)^2}{4}\tilde{b}_2,
\ena
we get the allowed regions of our parameters $(\la,\c)$ for the existence of
asymptotically AdS solutions. Because these are the same as in our previous
paper,~\cite{GOT2} we refer the reader to that paper for the explicit regions.

%%%%%%%%%%%%%%%%%%%%%%%%%%%%%%%%%%%%
%%%%%%%%%%%%%%%%%%%%%%%%%%%%%%%%%%%%
\section{Spherically symmetric black hole solutions with $k=1$}
\label{spherical}
%%%%%%%%%%%%%%%%%%%%%%%%%%%%%%%%%%%%
%%%%%%%%%%%%%%%%%%%%%%%%%%%%%%%%%%%%

We construct our black hole solutions numerically for the parameters
\bea
\c = \frac12,~~
\la = \frac13,~~
\tilde\Lambda=-1,~~
\label{bc1}
\ena
and the conditions
\bea
\phi_-=0,~~
\d_0=0,
\label{bc2}
\ena
in $D=4$ -- 6 and 10.

We integrate the field equations from the event horizon to infinity.
The first step in the procedure is to choose appropriate radius of the event
horizon~$\tilde{r}_H$. We find that we should choose large radius in this process,
for instance $\tilde{r}_H\sim 100$, because there are cases where no solution exists for
small horizon radii. Next we choose the value of $\phi_H$ and determine the values
of other fields by Eqs.~\p{bhor} and \p{pder}.
Since the condition \p{bc2} is in general not satisfied for most of the values
$\phi_H$, it should be tuned such that $\phi_-=0$ is achieved in the asymptotic
behavior~\p{behavior}.
We also fix $\delta_H=0$ in the integration and this would give nonzero $\d_0$.
However $\delta_0=0$ is always realized by the shift symmetry \p{sym3}
so we do not have to worry about that.
As a result, there is only one freedom of choosing $\tilde{r}_H$,
given a cosmological constant.
The solutions are obtained for the particular choice of $\gamma$ and $\lambda$,
but we expect that qualitative properties do not change for other choices of
these parameters, though there is an indication that the range of the horizon radii
for the existence of the black hole solutions changes depending on the strength
of the dilaton coupling $\c$.~\cite{MOS}
Using the symmetry in Eq.~\p{sym2}, the solutions for different values of
cosmological constant can be generated from a solution for $\tilde\Lambda=-1$.
Indeed, solutions for $\tilde{\Lambda}_1$ can be obtained by simply changing
the variables as
\begin{equation}
\tilde{r} \to |\tilde{\Lambda}_1|^{-\frac{\gamma}{2(\lambda-\gamma)}}\tilde{r},
~~~~~
\phi \to \phi-\frac{1}{\lambda-\gamma}\log|\tilde{\Lambda}_1|,
~~~~~
\tilde{m}_g \to |\tilde{\Lambda}_1|^{-\frac{(D-3)\gamma}{2(\lambda-\gamma)}}\tilde{m}_g.
\end{equation}

\begin{figure}[t]
\begin{center}
\includegraphics[width=4cm]{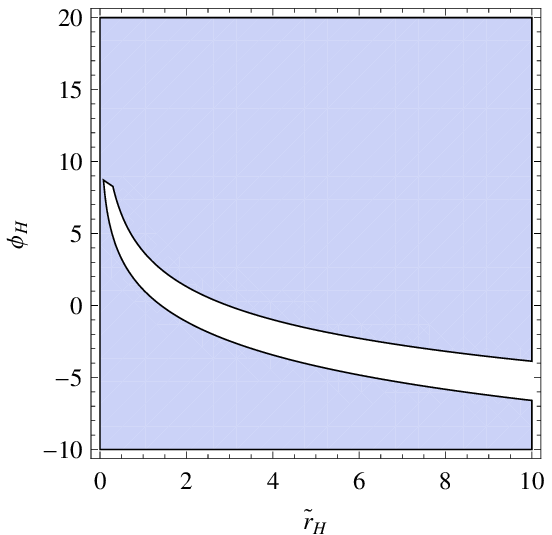}~~
\put(-60,-15){(a)}
%\hspace{2mm}
\includegraphics[width=4cm]{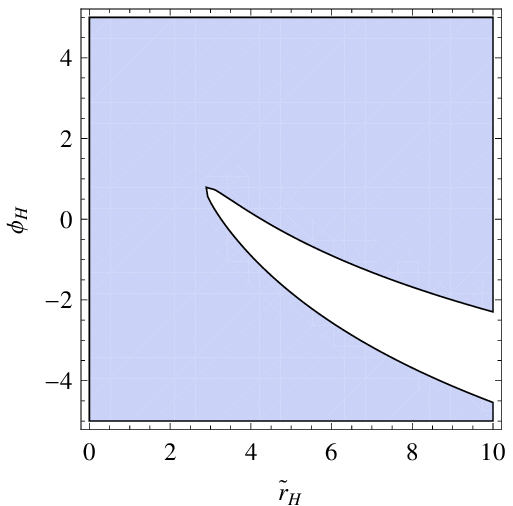}~~
\put(-60,-15){(b)}
%\vspace{2mm}
\includegraphics[width=4cm]{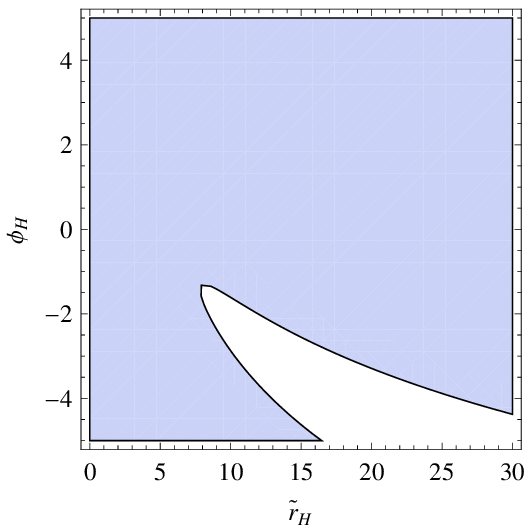}~~
\put(-60,-15){(c)}
%\hspace{2mm}
\includegraphics[width=4cm]{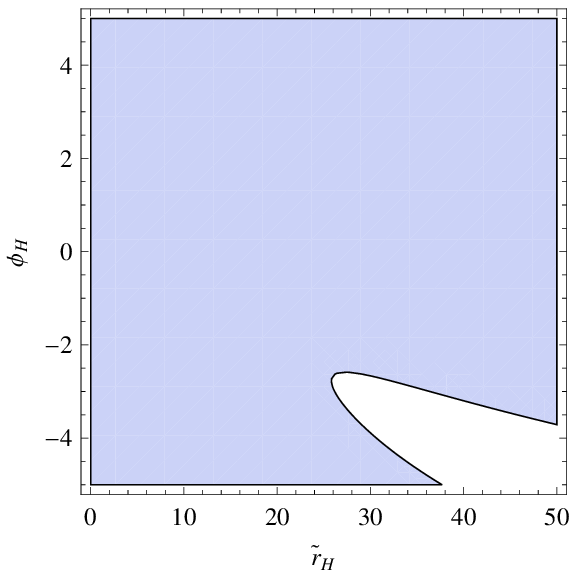}
\put(-60,-15){(d)}
\end{center}
\caption{
The regions where Eq. (\ref{pder}) has real roots of $\phi_H'$ (shaded area)
for $k=1$ and $\tilde{\Lambda}=-1$ in (a) $D=4$,  (b) $D=5$,  (c) $D=6$,  and (d) $D=10$.
On the boundary $\phi_H'$  degenerates. Only in these regions, regularity at the event
horizon can be satisfied.
}
\label{f1}
\end{figure}

Let us examine the boundary condition \p{pder} in more detail here.
It is a quadratic equation with respect to $\phi_H'$. To guarantee the reality of
$\phi_H'$, we find that there are forbidden parameter regions for $\tr_H$ and $\phi_H$.
Such regions (or allowed regions) are depicted in Fig.~\ref{f1} for our choice of
parameters $\gamma$, $\lambda$ and $\tilde\Lambda$ in each dimension.
The solutions can exist only in those shaded regions.
However, this does not mean that there are always solutions for those values
in these regions. We also mention that the regions seem to change depending on
the parameter choice of $\c$ and $\la$,~\cite{MOS} but we will not discuss
this issue in this paper.

% figures ------------------------------------------------------------------------
% figures --------------------------------------------------------------------------

In the following subsections, numerical solutions are presented in various dimensions.
Before doing that, let us consider the analytical solution. It can be confirmed that
the basic equations have the exact solution
\bea
\phi \equiv \phi_0,~~
\delta \equiv 0,~~
B = \tilde{b}_2 \tr^2+1,
\label{AdS}
\ena
where the parameters $\tilde{b}_2$ and $\phi_0$ are given in Eqs.~\p{b2inf}
and \p{phiinf2}.
This is the AdS solution and the spacetime is regular everywhere. To our knowledge,
this is the first example of the exact solution in the dilatonic Einstein-GB system.
It is important to note that there is not such a kind of exact solution without
cosmological constant. The effective potential of the dilaton field is composed of
two parts, the GB term and the cosmological constant. This allows the dilaton field
to have equilibrium point in the effective potential. Without cosmological constant,
however, the effective potential has slopes everywhere, and $\phi$ cannot be constant.

%%%%%%%%%%%%%%%%%%%%%%%%%%%%%%%%%%%%
\subsection{$D=4$ solution}
%%%%%%%%%%%%%%%%%%%%%%%%%%%%%%%%%%%%

\begin{figure}[h]
\begin{center}
\includegraphics[width=15cm]{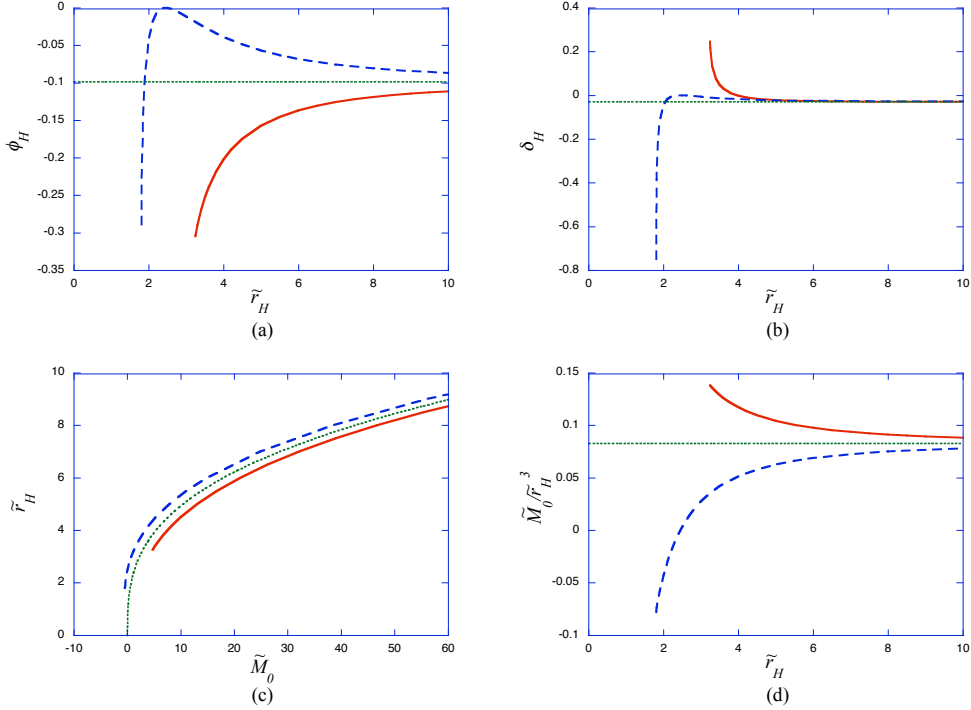}
\end{center}
\vspace{-5mm}
\caption{
Numerical values of the fields (a) $\phi_H$ and (b) $\delta_H$ at the horizon
and the mass of the black hole (c) $\tilde{M}_0$-$\tr_H$ diagram in $D=4$.
(d) shows the scaled mass by horizon radius. For each figures,
$k=1$ is in the solid (red) line, $k=0$ the dotted (green) line, and  $k=-1$
the dashed (blue) line.
The AdS radius is $\tilde{\ell}_{\rm AdS}= \sqrt{6}$, and the asymptotic value of
the dilaton is $\phi_0=0$.
}
\label{d4_parameter}
\end{figure}

First we present the black hole solutions for $D=4$. From Eqs.~\p{b2inf} and \p{phiinf2},
we see that the square of inverse AdS radius and the asymptotic value of the dilaton
field are $\tilde{b}_2=1/6$ ($\tilde{\ell}_{AdS}=\sqrt{6}$) and $\phi_0=0$,
respectively.
We show $\phi_H, \d_H$ and $\tilde{M}_0$ as functions of $\tr_H$ in
solid lines in Fig.~\ref{d4_parameter}.
$\phi_H$ is negative and less than that of the $k=0$ case for any horizon radius, and
approaches the value $\phi_H^{(k=0)}=-0.0985664$ in the $k=0$ case for large horizon
radius (Fig.~\ref{d4_parameter}(a)).
There is a forbidden region in the left side of the $\phi_H$-$\tr_H$ diagram
in Fig.~\ref{f1}(a). Consequently, as the horizon radius becomes small, the value
($\tr_H$, $\phi_H)$ hits the forbidden region, and the solution disappears
for horizon radius smaller than the critical value $\tilde{r}_H=3.245$.
For this critical horizon radius, the
second derivatives of the dilaton field $\phi''$ and $\delta'$ diverge at
the horizon while $B'_H$ and $\delta_H$ are finite (Fig.~\ref{d4_parameter}(b)).
This behavior is similar to the zero cosmological constant case discussed
in Refs.~\citen{TYM,GOT1}. Since the critical horizon radius is larger than the AdS
radius $\tilde{\ell}_{\rm AdS}=\sqrt{6}$, the horizon of every solution stays
outside AdS radius. The gravitational mass $\tilde{M}_0$ is monotonic with respect
to $\tr_H$ (Fig.~\ref{d4_parameter}(c)).
In the $k=0$ case,~\cite{GOT2} $\tilde{M}_0$ was proportional to
$\tr_H^{D-1}$ by the scaling symmetry. In the present case, it is found that
similar proportional relation is gradually realized for the large black holes
(Fig.~\ref{d4_parameter}(d)).

% figures --------------------------------------------------------------------------
% figures ---------------------------------------------------------------------------

% figures -------------------------------------------------------------------
\begin{figure}[h]
\begin{center}
\includegraphics[width=18cm]{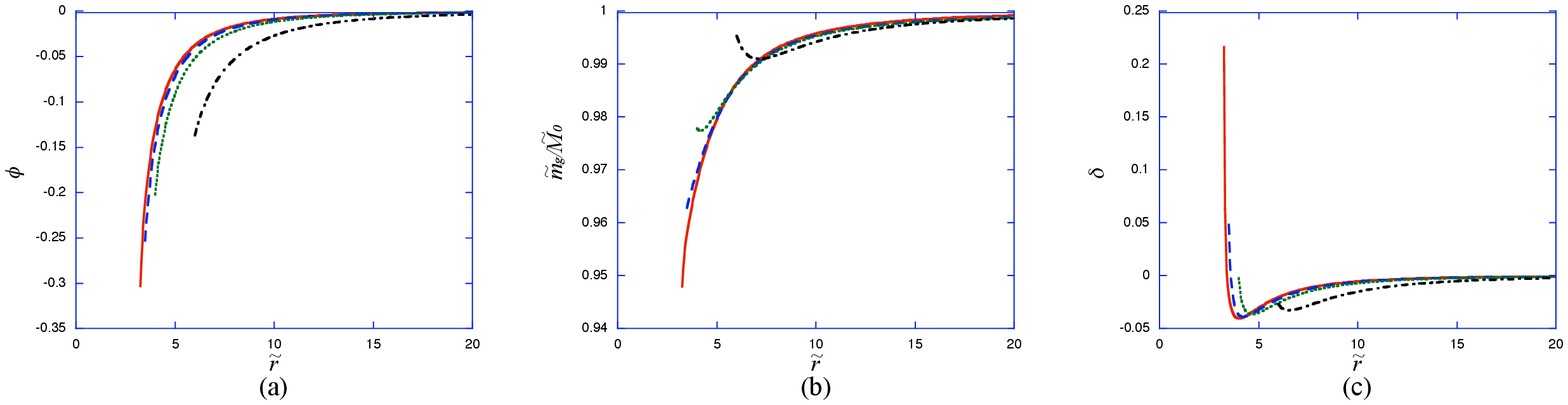}
%\put(-110,-20){(a)}
%\put(125,-20){(b)}
\end{center}
\vspace{-2mm}
\caption{
Behaviors of (a) the dilaton field $\phi(\tr)$, (b) the mass function
$\tm_g(\tr)/\tilde{M}_0$ and (c) the lapse function $\delta(\tr)$ of the black hole
solutions with $k=1$ in $D=4$. The horizon radii and the masses are
$\tr_H =3.25$, $\tilde{M}_0=4.7322$ (solid line),
$\tr_H =3.5$, $\tilde{M}_0=5.5284$ (dashed line),
$\tr_H=4.0$, $\tilde{M}_0=7.4995$ (dotted line),
and  $\tr_H=6.0$, $\tilde{M}_0=21.098$ (dot-dashed line).
}
\label{d4_config}
\end{figure}
% figures ---------------------------------------------------------------------

The configurations of the fields in the black hole solutions
are depicted in Fig.~\ref{d4_config}.
The dilaton field increases monotonically to its asymptotic value $\phi_0=0$.
The mass function increases monotonically for smaller black holes
while it decreases near the event horizon for larger black holes.
As the horizon radius approaches the critical value, the function $\delta$
is finite but becomes steep around the event horizon.

%%%%%%%%%%%%%%%%%%%%%%%%%%%%%%%%%%%%
\subsection{$D=5$ solution}
%%%%%%%%%%%%%%%%%%%%%%%%%%%%%%%%%%%%

\begin{figure}[ht]
\begin{center}
\includegraphics[width=15cm]{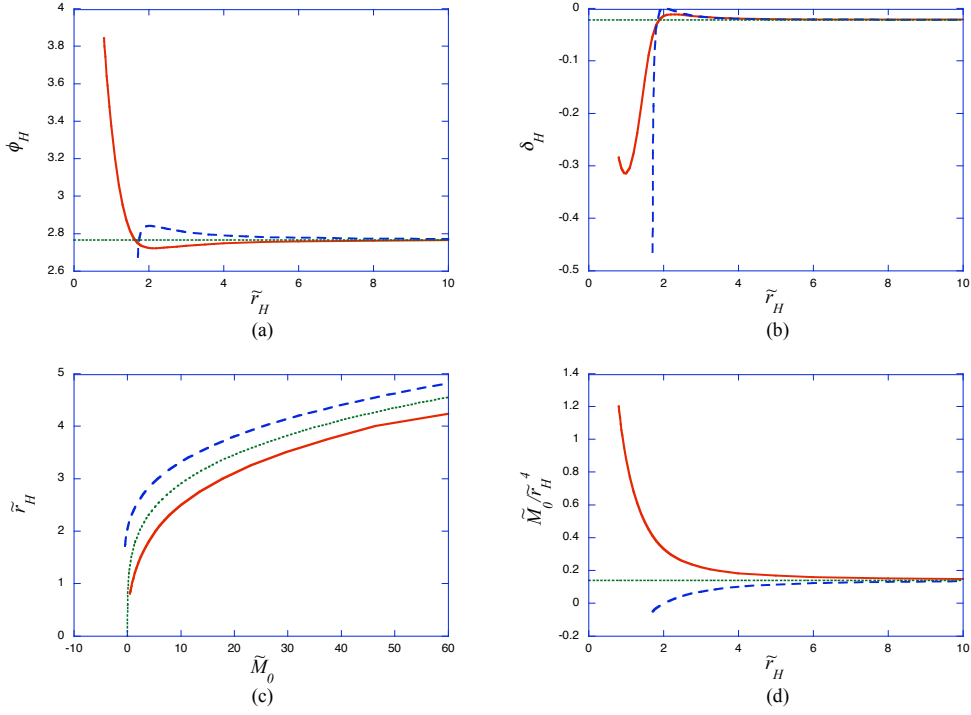}
%\put(-110,-20){(a)}
%\put(125,-20){(b)}
\end{center}
\vspace*{-5mm}
\caption{
Numerical values of the fields (a) $\phi_H$ and (b) $\delta_H$ at the horizon,
and the mass of the black hole (c) $\tilde{M}_0$-$\tr_H$ diagram in $D=5$.
(d) shows the scaled mass by horizon radius. For each figure,
$k=1$ is in the solid (red) line, $k=0$ the dotted (green) line, and  $k=-1$
the dashed (blue) line.
The AdS radius is $\tilde{\ell}_{\rm AdS}=2.0267$, and the asymptotic value of
the dilaton is $\phi_0=2.84082$.
}
\label{d5_parameter}
\end{figure}

It follows from Eqs.~\p{b2inf} and \p{phiinf2} that
the square of inverse AdS radius and the asymptotic value of the dilaton
field are $\tilde{b}_2=0.24346$ ($\tilde{\ell}_{\rm AdS}=2.0267$) and $\phi_0=2.84082$,
respectively.
We show $\phi_H, \d_H$ and $\tilde{M}_0$ as functions of $\tr_H$ in
Fig.~\ref{d5_parameter}.
For the large black hole (naively for $\tr_H\gtrsim \tilde{\ell}_{\rm AdS}$),
$\phi_H$ is smaller than $\phi_0$ as in the $D=4$ case (Fig.~\ref{d5_parameter}(a)).
$\phi_H$ approaches $\phi_H^{(k=0)}=2.767015$ in the $\tr_H\to \infty$ limit.
However, for the small black holes, $\phi_H$ is larger than $\phi_0$.
There is a forbidden region in the $\phi_H$-$\tr_H$ diagram in Fig.~\ref{f1}(b),
and as the  horizon radius becomes smaller, black hole solution disappears as
in $D=4$. However, in $D=5$, the disappearance occurs before
the parameters ($\tr_H$, $\phi_H)$ hit the forbidden region in Fig.~\ref{f1}(b).
For the critical horizon radius $\tilde{r}_H=0.805$, below which there is no black
hole solution, the horizon is regular but $\phi''$ diverges at $\tr=\tr_s\equiv 1.11896$.
By the field equations,
the derivative of the metric functions $B'$ and $\delta'$ also diverge.
The Kretschmann invariant
\begin{eqnarray}
&&
R^{\mu\nu\rho\sigma}R_{\mu\nu\rho\sigma}
= (B''-3B'\d'+2B(\d'^2-\d''))^2 +\frac{2(D-2)}{r^2} (B'^2-2BB'\d'+2B^2 \d'^2)
\nonumber \\
&& \hspace{30mm}
+ \frac{2(D-2)_3}{r^4} (k-B)^2,
\end{eqnarray}
diverges there, and hence the surface $\tr=\tr_s$ is curvature singularity.
These properties are quite different from the zero cosmological constant case,~\cite{GOT1}
where the black hole solution exists for any horizon radius.
It is possible, however, that these behaviors change if we change the parameter
$\c$.~\cite{MOS}
It is interesting that the cosmological constant changes the small scale
structure compared to the bare curvature radius defined by
$\tilde{\ell}^2=-(D-1)_2/\tilde\Lambda$, although it usually affects the properties
of large structure.
Relation between $\tilde{M}_0$ and $\tr_H$ is monotonic  (Fig.~\ref{d5_parameter}(c)).
It is also found that for the large horizon radius $\tilde{M}_0$ is proportional
to $\tr_H^{\:4}$ (Fig.~\ref{d5_parameter}(d)).
% figures ------------------------------------------------------------------------
% figures -------------------------------------------------------------------------

\begin{figure}[ht]
\begin{center}
\includegraphics[width=18cm]{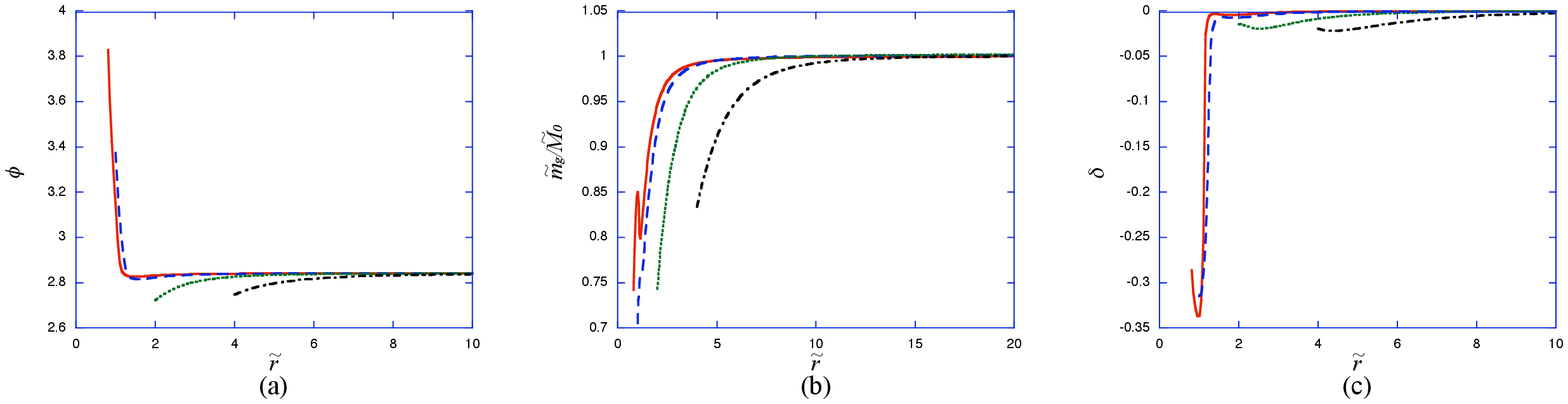}
%\put(-110,-20){(a)}
%\put(125,-20){(b)}
\end{center}
\vspace{-2mm}
\caption{
Behaviors of (a) the dilaton field $\phi(\tr)$, (b) the mass function
$\tm_g(\tr)/\tilde{M}_0$ and (c) the lapse function $\delta(\tr)$ of the black hole
solutions with $k=1$ in $D=5$. The horizon radii and the masses are
$\tr_H =0.81$, $\tilde{M}_0=0.51273$ (solid line),
$\tr_H =1.0$, $\tilde{M}_0=0.88107$ (dashed line),
$\tr_H=2.0$, $\tilde{M}_0=5.3094$ (dotted line),
and  $\tr_H=4.0$, $\tilde{M}_0=46.417$ (dot-dashed line).
}
\label{d5_config}
\end{figure}

We show the configurations of the fields in the black hole solutions
in Fig.~\ref{d5_config}.
For large black holes, the dilaton field increases monotonically to
its asymptotic value $\phi_0=2.84082$. The mass function also increases
monotonically. On the other hand, for small black holes, the dilaton
field decreases rapidly around the event horizon and becomes smaller than
$\phi_0$. Around these decreasing regions, the mass function varies violently.
The dilaton field increases outside some radius towards its
asymptotic value.
% figures ------------------------------------------------------------
% figures ---------------------------------------------------------------------------

%%%%%%%%%%%%%%%%%%%%%%%%%%%%%%%%%%%%
\subsection{$D=6$ solution}
%%%%%%%%%%%%%%%%%%%%%%%%%%%%%%%%%%%%

\begin{figure}[ht]
\begin{center}
\includegraphics[width=15cm]{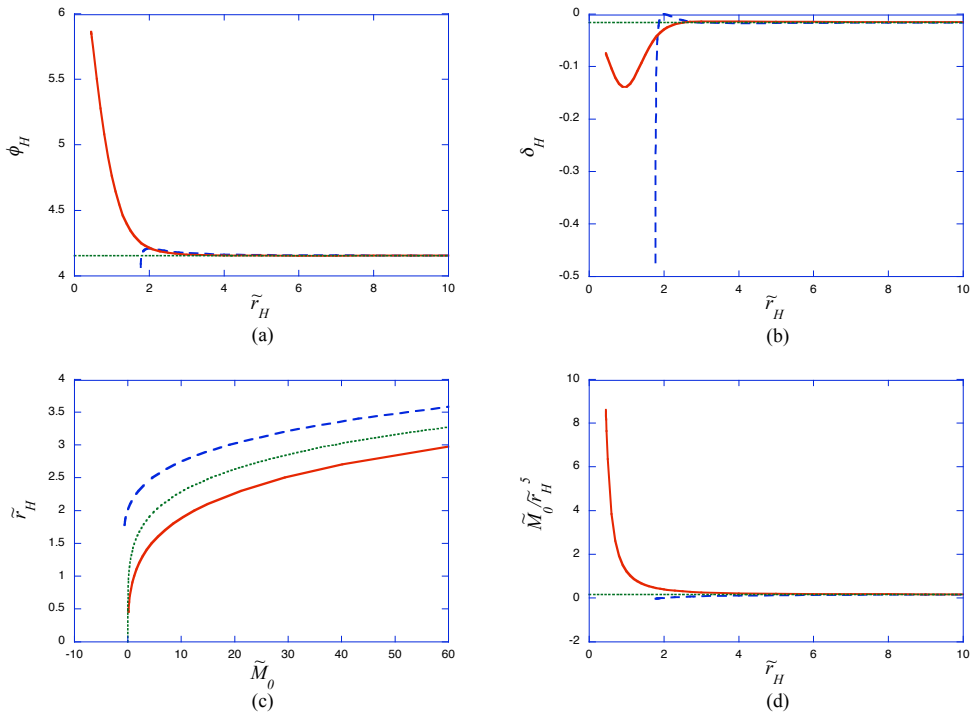}
%\put(-110,-20){(a)}
%\put(125,-20){(b)}
\end{center}
\vspace{-5mm}
\caption{Numerical values of the fields (a) $\phi_H$ and (b) $\delta_H$ at the horizon
and the mass of the black hole (c) $\tilde{M}_0$-$\tr_H$ diagram in $D=6$.
(d) shows the scaled mass by horizon radius. For each figure,
$k=1$ is in the solid (red) line, $k=0$ the dotted (green) line, and  $k=-1$
the dashed (blue) line.
The AdS radius is $\tilde{\ell}_{\rm AdS}=2.0059$, and the asymptotic value of
the dilaton is $\phi_0=4.20868$.
}
\label{d6_parameter}
\end{figure}

We see from Eqs.~\p{b2inf} and \p{phiinf2} that
the square of inverse AdS radius and the asymptotic value of the dilaton
field are $\tilde{b}_2=0.248535$ ($\tilde{\ell}_{\rm AdS}=2.0059$) and $\phi_0=4.20868$,
respectively.
We show $\phi_H, \d_H$ and $\tilde{M}_0$ as functions of $\tr_H$ in solid lines in
Fig.~\ref{d6_parameter}.
The qualitative properties are the same as those in the $D=5$ case.
For the large black hole ($\tr_H\gtrsim \tilde{\ell}_{\rm AdS}$), $\phi_H$ is smaller
than $\phi_0$ (Fig.~\ref{d6_parameter}(a)).
$\phi_H$ approaches  $\phi_H^{(k=0)}=4.15418$ in the $\tr_H\to \infty$ limit.
However, for the small black holes, $\phi_H$ is larger than $\phi_0$.
As the horizon radius becomes smaller, the black hole solution disappears
at the critical radius $\tilde{r}_H=0.805$.
As is seen from Fig.~\ref{f1}(c), a forbidden region exists far below in
the $\phi_H$-$\tr_H$ diagram. Hence the disappearance of
solution is not due to hitting the forbidden region but due to diverging behavior
of derivatives of the fields outside the event horizon as in $D=5$.
The gravitational mass $\tilde{M}_0$ is monotonic with respect to $\tr_H$
(Fig.~\ref{d6_parameter}(c)).
It is found that for the large horizon radius $\tilde{M}_0$ is proportional
to $\tr_H^{\:5}$ (Fig.~\ref{d6_parameter}(d)).
\vspace{-5mm}
% figures ----------------------------------------------------------------------------
% figures ----------------------------------------------------------------------------
% figures ------------------------------------------------------------------------------
\begin{figure}[ht]
\begin{center}
\includegraphics[width=18cm]{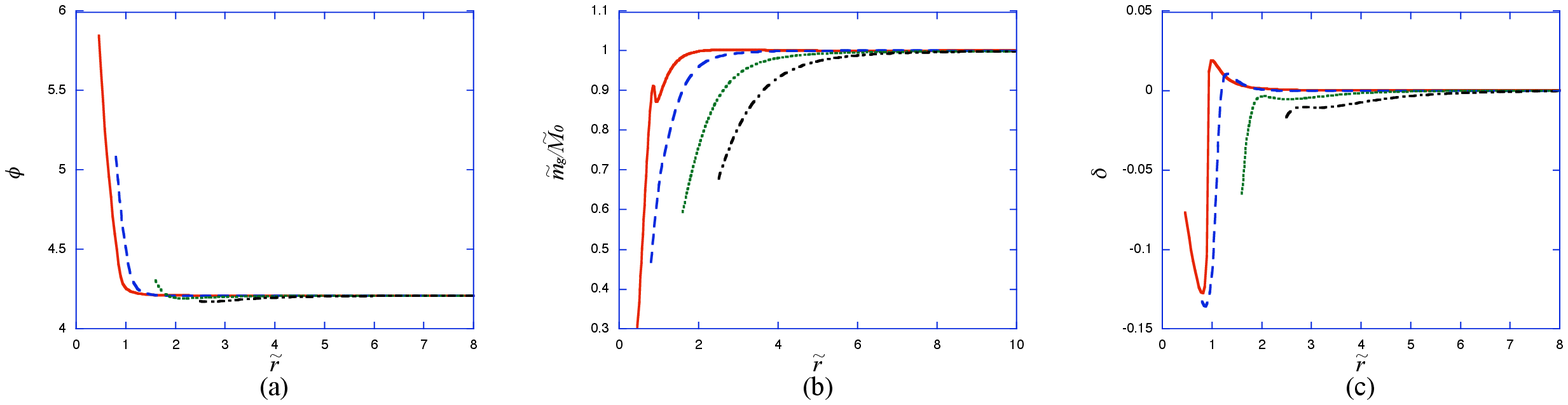}
%\put(-110,-20){(a)}
%\put(125,-20){(b)}
\end{center}
\vspace{-2mm}
\caption{
Behaviors of (a) the dilaton field $\phi(\tr)$, (b) the mass function
$\tm_g(\tr)/\tilde{M}_0$ and (c) the lapse function $\delta(\tr)$ of the black
hole solutions with $k=1$ in $D=6$. The horizon radii and the masses are
$\tr_H =0.46$, $\tilde{M}_0=0.16813$ (solid line),
$\tr_H =0.8$, $\tilde{M}_0=0.63364$ (dashed line),
$\tr_H=1.6$, $\tilde{M}_0=5.6443$ (dotted line),
and  $\tr_H=2.5$, $\tilde{M}_0=29.449$ (dot-dashed line).
}
\label{d6_config}
\end{figure}
% figures --------------------------------------------------------------------------

We show the configurations of the fields in the black hole solutions
in Fig.~\ref{d6_config}.
For large black holes, the dilaton field increases monotonically to its asymptotic
value $\phi_0=4.20868$. The mass function also increases monotonically.
On the other hand, for small black holes, the dilaton field decreases around
the event horizon and increases towards its asymptotic value.
There is a region where the mass function decreases.

%%%%%%%%%%%%%%%%%%%%%%%%%%%%%%%%%%%%
\subsection{$D=10$ solution}
%%%%%%%%%%%%%%%%%%%%%%%%%%%%%%%%%%%%

\begin{figure}[htb]
\begin{center}
\includegraphics[width=15cm]{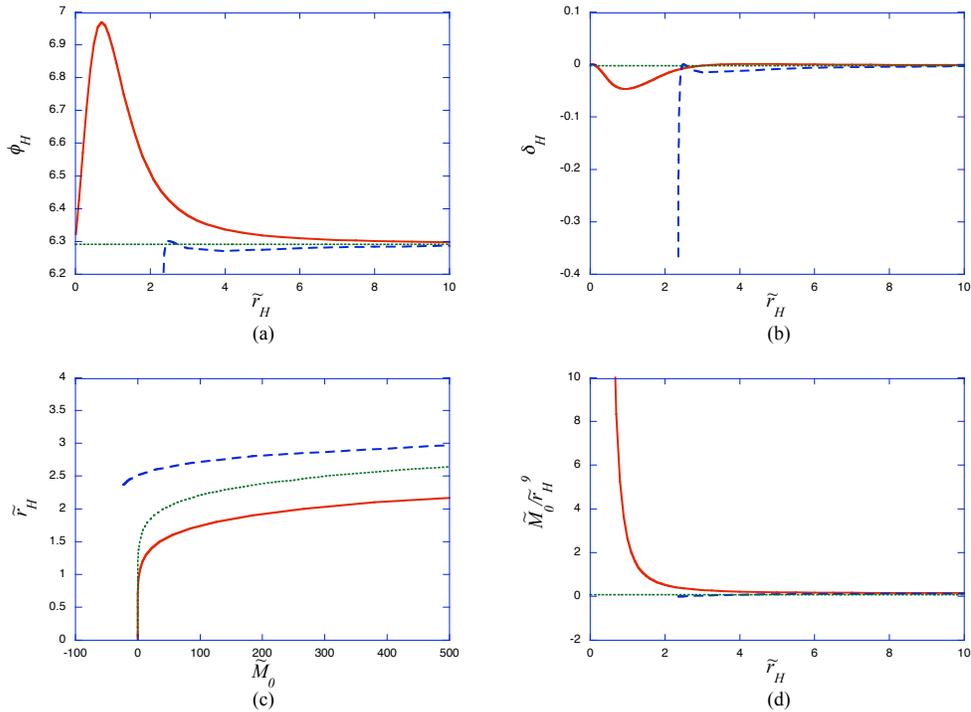}
%\put(-110,-20){(a)}
%\put(125,-20){(b)}
\end{center}
\vspace{-5mm}
\caption{Numerical values of the fields (a) $\phi_H$ and (b) $\delta_H$ at the horizon
and the mass of the black hole (c) $\tilde{M}_0$-$\tr_H$ diagram in $D=10$.
(d) shows the scaled mass by horizon radius. For each figure,
$k=1$ is in the solid (red) line, $k=0$ the dotted (green) line, and  $k=-1$
the dashed (blue) line.
The AdS radius is $\tilde{\ell}_{\rm AdS}= 2.5089$, and the asymptotic value of
the dilaton is $\phi_0=6.30143$.
}
\label{d10_parameter}
\end{figure}

It follows from Eqs.~\p{b2inf} and \p{phiinf2} that
the square of inverse AdS radius and the asymptotic value of the dilaton
field are $\tilde{b}_2=0.158862$ ($\tilde{\ell}_{\rm AdS}=2.5089$) and $\phi_0=6.30143$,
respectively.
We show $\phi_H, \d_H$ and $\tilde{M}_0$ as functions of $\tr_H$ in solid lines in
Fig.~\ref{d10_parameter}.
% figures ------------------------------------------------------------------------------
% figures ---------------------------------------------------------------------------
% figures ---------------------------------------------------------------------------
\begin{figure}[h]
%\begin{center}
\includegraphics[width=18cm]{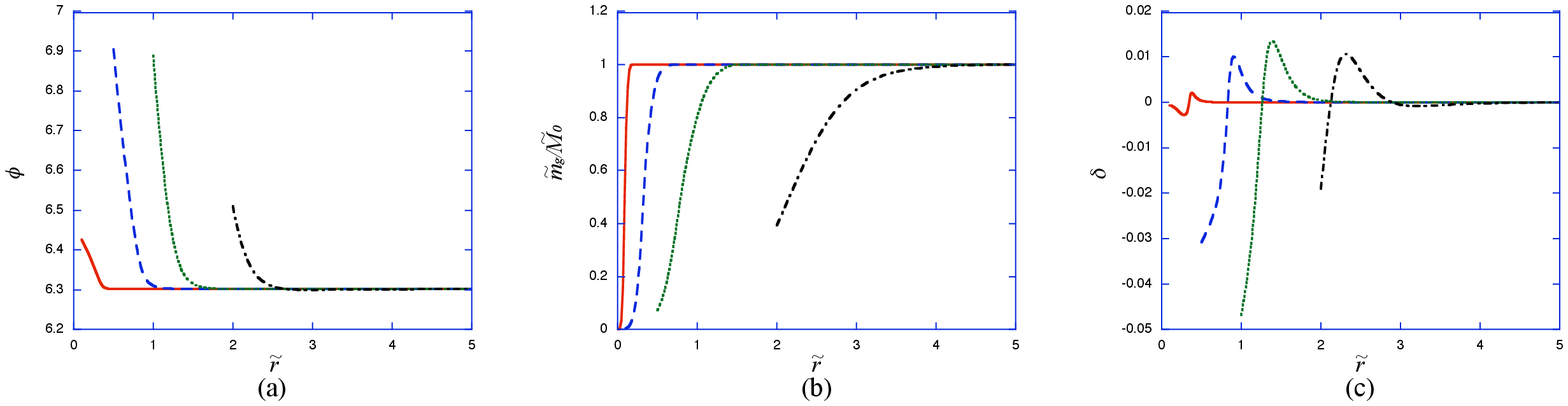}
%\put(-110,-20){(a)}
%\put(125,-20){(b)}
%\end{center}
\vspace{-2mm}
\caption{
Behaviors of (a) the dilaton field $\phi(\tr)$, (b) the mass function
$\tm_g(\tr)/\tilde{M}_0$ and (c) the lapse function $\delta(\tr)$ of the black hole
solutions with $k=1$ in $D=10$. The horizon radii and the masses are
$\tr_H =0.01$, $\tilde{M}_0=2.0898\times 10^{-10}$ (solid line),
$\tr_H =0.1$, $\tilde{M}_0=1.9999\times 10^{-5}$ (dashed line),
$\tr_H=0.5$, $\tilde{M}_0=5.5699\times 10^{-2}$ (dotted line),
and  $\tr_H=2.0$, $\tilde{M}_0=2.6667\times 10^{2}$ (dot-dashed line).
}
\label{d10_config}
\end{figure}
% figures -----------------------------------------------------------------------------
For the large black hole ($\tr_H\gtrsim 8$), $\phi_H$ is smaller than $\phi_0$
(Fig.~\ref{d10_parameter}(a)).
$\phi_H$ approaches $\phi_H^{(k=0)}=6.2916147$ in the $\tr_H\to \infty$ limit.
For the small black holes, $\phi_H$ is larger than $\phi_0$.
These qualitative properties are the same as those in $D=5$ and 6 case.
However, as the horizon radius becomes smaller, $\phi_H$ approaches $\phi_0$,
and the solution continues to exist to zero horizon limit. This means that
there is no critical horizon radius, and black hole solution exists for
any horizon radius according to the numerical analysis.
As is seen from  Fig.~\ref{f1}(d), a forbidden region exists far below in the
$\phi_H$-$\tr_H$ diagram. As the horizon radius
becomes zero, variables $\phi'$, $\phi''$, and $\delta_H$ becomes zero.
Hence the solution is connected to the regular AdS solution \p{AdS}.\footnote{
This does not mean that the black hole solution changes to the regular AdS solution
continuously through, for example, Hawking radiation.}
This is similar to the zero cosmological constant case\cite{GOT1}.
Relation between $\tilde{M}_0$ and $\tr_H$ is monotonic  (Fig.~\ref{d10_parameter}(c)).
It is found that for the large horizon radius $\tilde{M}_0$ is proportional
to $\tr_H^{\:9}$ (Fig.~\ref{d10_parameter}(d)).
On the other hand, $\tilde{M}_0$ is proportional to $\tr_H^{\:5}$ for the small
horizon radius. These relations are the same as the non-dilatonic case.

We show the configurations of the fields in the black hole solutions
in Fig.~\ref{d10_config}.
For large black holes ($\tr_H \gtrsim 8$), the dilaton field increases monotonically
to its asymptotic value $\phi_0=6.30143$. The mass function also increases monotonically.
On the other hand, for small black holes, the dilaton field decreases around
the event horizon and increases towards its asymptotic value. The mass function
also increases monotonically.

%%%%%%%%%%%%%%%%%%%%%%%%%%%%%%%%%%%%
%%%%%%%%%%%%%%%%%%%%%%%%%%%%%%%%%%%%
\section{Hyperbolic black holes with $k=-1$}
\label{hyperbolic}
%%%%%%%%%%%%%%%%%%%%%%%%%%%%%%%%%%%%
%%%%%%%%%%%%%%%%%%%%%%%%%%%%%%%%%%%%

The procedure to calculate the black hole solutions are the same as that in $k=1$ case.
It can be confirmed that the basic equations have the exact solution
\bea
\phi \equiv \phi_0,~~
\delta \equiv 0,~~
B = \tilde{b}_2 \tr^2-1.
\label{AdSBH}
\ena
where the parameters $b_2$ and $\phi_0$ are again given in Eqs.~\p{b2inf} and \p{phiinf2}.
This solution has an event horizon at $\tr=1/\sqrt{\tilde{b}_2}$, which coincides
with the AdS radius $\tilde \ell_{\rm AdS}$. It follows from Eq.~(\ref{mass}) that
the mass of the black hole solution is $\tm_g\equiv \tilde{M}_0 =0$.
Hence this solution is the zero mass black hole. The center of the solution
is not singular but regular.

\begin{figure}[h]
\begin{center}
\includegraphics[width=4cm]{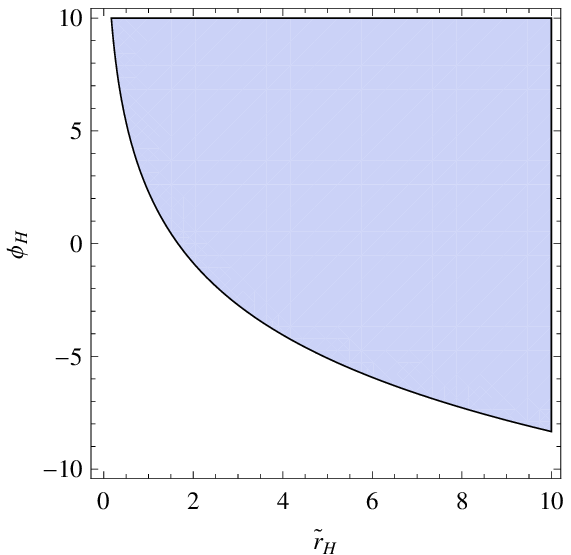}~~
\put(-68,-15){(a)}
%\hspace{10mm}
\includegraphics[width=4cm]{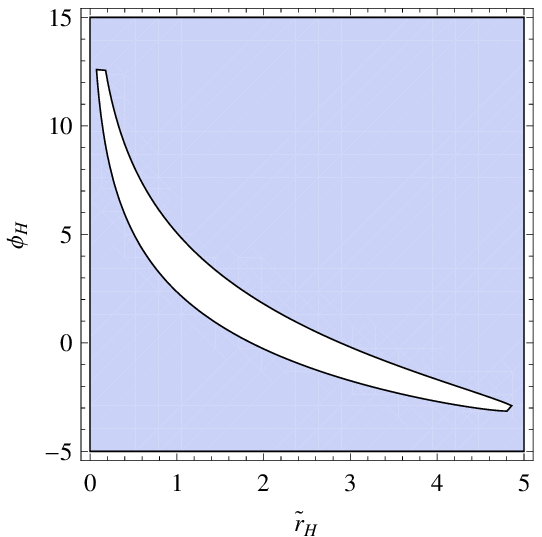}~~
\put(-68,-15){(b)}
%\vspace{6mm}
\includegraphics[width=4cm]{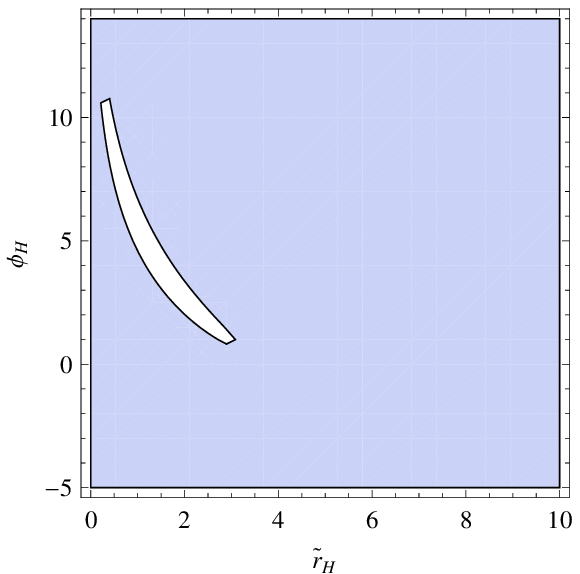}~~
\put(-68,-15){(c)}
%\hspace{10mm}
\includegraphics[width=4cm]{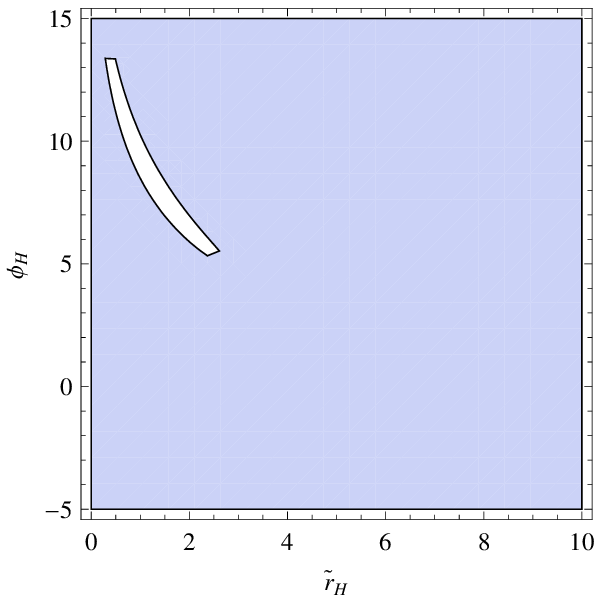}
\put(-68,-15){(d)}
\end{center}
\caption{The regions where Eq. (\ref{pder}) has real roots of $\phi_H'$ (shaded area)
for $k=-1$ and $\tilde{\Lambda}=-1$ in (a) $D=4$,  (b) $D=5$,  (c) $D=6$,  and (d) $D=10$.
On the boundary $\phi_H'$  degenerates. Only in these regions, regularity at the event
horizon can be satisfied.
}
\label{f5}
\end{figure}

The allowed parameter regions of ($\phi_H'$, $\tr$) are shown in Figs.~\ref{f5}.
Note that the forbidden regions appear in the larger values of $\phi_H$ in contrast
to $k=1$ case.

% figures ------------------------------------------------------------------------
% figures ---------------------------------------------------------------------------

It turns out that qualitative properties of the black hole solutions do not depend on
the spacetime dimensions. So, we present the solutions in $D=4$ -- 6, and 10 together.

The asymptotic value of the dilaton field $\phi_0$ and the square of inverse AdS radius
$\tilde{b}_2=\tilde{\ell}_{\rm AdS}^{-2}$ are the same as those in the $k=1$ case for
any dimensions.
We show $\phi_H$, $\d_H$ and $\tilde{M}_0$ as functions of $\tr_H$ in
Figs.~\ref{d4_parameter}, \ref{d5_parameter}, \ref{d6_parameter} and \ref{d10_parameter}
for $D=4$, 5, 6 and $10$, respectively, all in dashed lines.
The solutions are classified by their horizon radius as $\tr_H>\tilde{\ell}_{\rm AdS}$ or
$\tr_H<\tilde{\ell}_{\rm AdS}$.
For the large black holes ($\tr_H>\tilde{\ell}_{\rm AdS}$), the dilaton field at
the horizon $\phi_H$ is larger than that in the $k=0$ case, which is opposite to
the $k=1$ case, but smaller than its asymptotic value $\phi_0$.
As the $\tr_H$ becomes large,  $\phi_H$ decreases, and approaches $\phi_H^{(k=0)}$
in the $\tr_H\to \infty$ limit.
The gravitational mass $\tilde{M}_0$ is monotonic with respect to $\tr_H$, and
proportional to $\tr_H^{D-1}$ for large $\tr_H$.
It is smaller than those of the $k=0$ and $+1$ with the same horizon radius.

The solution with $\tr_H=\tilde{\ell}_{\rm AdS}$ is given by Eq.~\p{AdSBH}, which
is the zero mass black hole.
In the non-dilatonic case, there are two branches of black hole solutions, i.e.,
the GR branch and the GB branch.
In each of them, there is a zero mass black hole solution. In the dilatonic case,
however, there is only a single zero mass solution, which corresponds to the GR branch.

For the solution with $\tr_H<\tilde{\ell}_{\rm AdS}$, $\phi_H$ decreases
as the horizon radius becomes small, and the values ($\tr_H$, $\phi_H$) hit
the forbidden region in Fig.~\ref{f5}.
There is no solution for smaller horizon radius than the critical radius
$\tr_H = 1.807$, $1.712$, $1.777$ and $2.364$, for $D=4$, 5, 6 and $10$, respectively.
For the critical horizon radius, the second derivative of the dilaton field $\phi''$
diverges at the horizon. These are similar to the $D=4$ and $k=1$ case.
In the non-dilatonic case, there is also a lower bound of the size of black hole
solution in the GR-branch. In that case, two horizons (the black hole event horizon
and the inner horizon) coincide, and the event horizons degenerate. In the present
dilatonic case, however, the horizon becomes singular for the critical radius.
The mass of these solution are negative.

\begin{figure}[ht]
\begin{center}
\includegraphics[width=18cm]{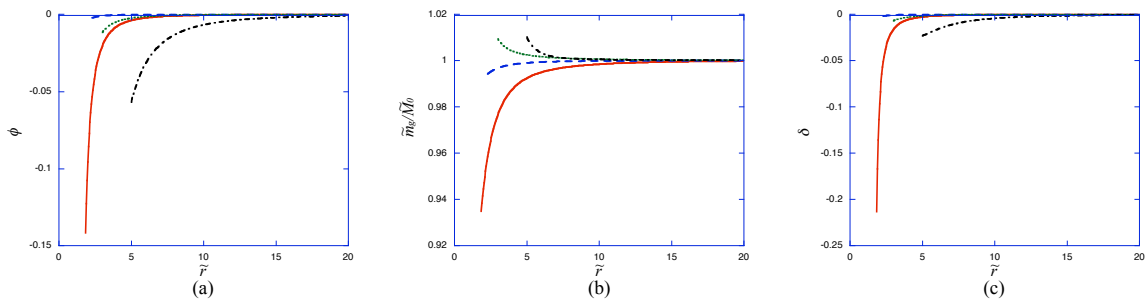}
%\put(-110,-20){(a)}
%\put(125,-20){(b)}
\end{center}
\vspace{-2mm}
\caption{Behaviors of (a) the dilaton field $\phi$, (b) the mass function
$\tm_g/\tilde{M}_0$ and (c) the lapse function $\delta$ of the black hole solutions
with $k=-1$
in $D=4$. The horizon radii  and the masses are
$\tr_H =1.85$, $\tilde{M}_0=-0.42504$  (solid line),
$\tr_H =2.3$, $\tilde{M}_0=-0.13686$  (dashed line),
$\tr_H=3.0$, $\tilde{M}_0=0.74302$  (dotted line),
and
$\tr_H=5.0$, $\tilde{M}_0=7.8371$ (dot-dashed line).
}
\label{d4-_config}
\end{figure}

\begin{figure}[ht]
\begin{center}
\includegraphics[width=18cm]{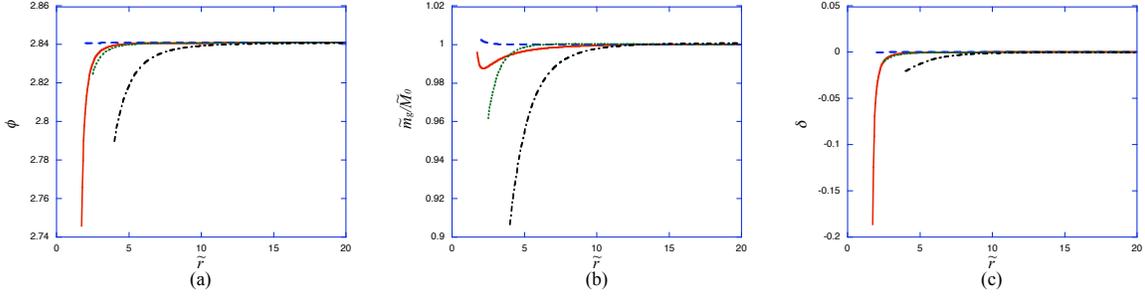}
%\put(-110,-20){(a)}
%\put(125,-20){(b)}
\end{center}
\vspace{-2mm}
\caption{Behaviors of (a) the dilaton field $\phi$, (b) the mass function
$\tm_g/\tilde{M}_0$ and (c) the lapse function $\delta$ of the black hole solutions
with $k=-1$ in $D=5$. The horizon radii  and the masses are
$\tr_H =1.73$, $\tilde{M}_0=-0.40774$ (solid line),
$\tr_H =2.0$, $\tilde{M}_0=-5.2196\times 10^{-2}$ (dashed line),
$\tr_H=2.5$, $\tilde{M}_0=1.6951$ (dotted line), and
$\tr_H=4.0$, $\tilde{M}_0=25.549$
(dot-dashed line).
}
\label{d5-_config}
\end{figure}

\begin{figure}[ht]
\begin{center}
\includegraphics[width=18cm]{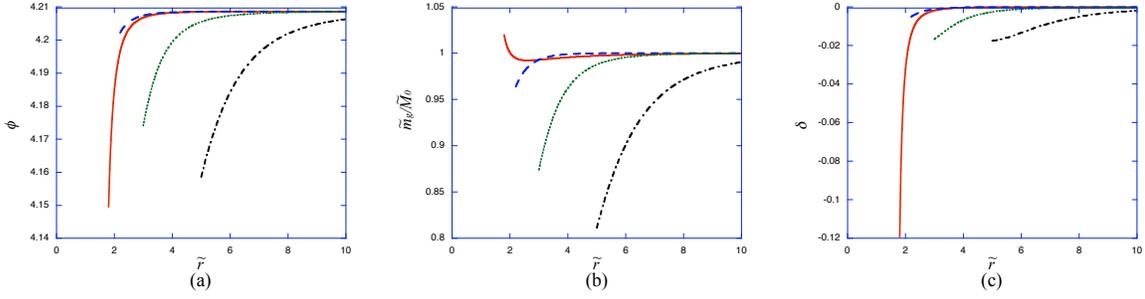}
%\put(-110,-20){(a)}
%\put(125,-20){(b)}
\end{center}
\vspace{-2mm}
\caption{Behaviors of (a) the dilaton field $\phi$, (b) the mass function
$\tm_g/\tilde{M}_0$ and (c) the lapse function $\delta$ of the black hole solutions
with $k=-1$ in $D=6$. The horizon radii  and the masses are
$\tr_H =1.82$, $\tilde{M}_0=-0.51938$ (solid line),
$\tr_H =2.2$, $\tilde{M}_0=1.1209$ (dashed line),
$\tr_H=3.0$, $\tilde{M}_0=19.097$ (dotted line), and
$\tr_H=5.0$, $\tilde{M}_0=4.0156\times 10^{2}$
(dot-dashed line).
}
\label{d6-_config}
\end{figure}

\begin{figure}[ht]
\begin{center}
\includegraphics[width=18cm]{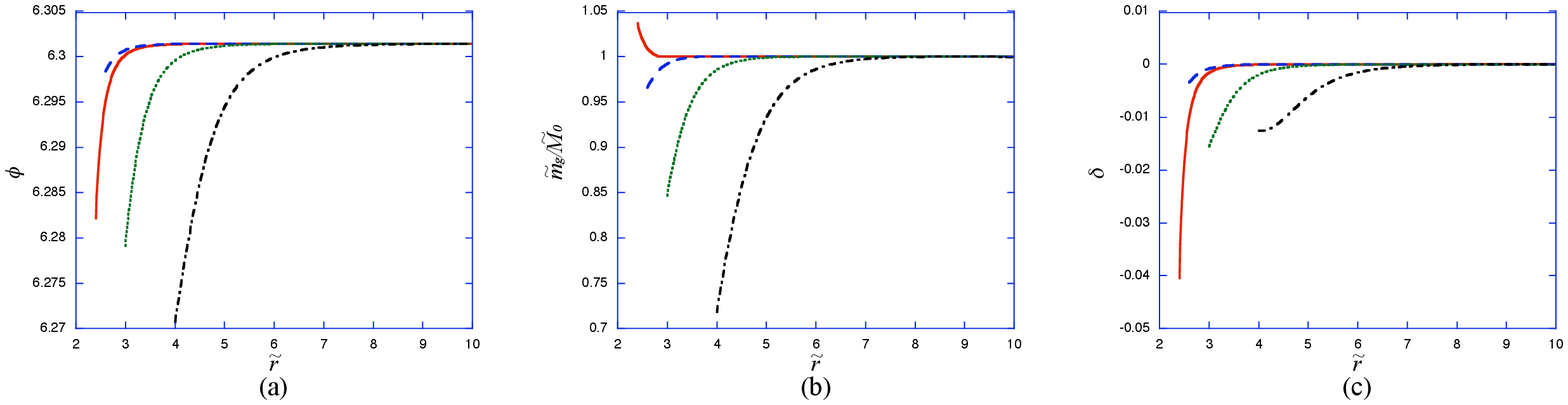}
%\put(-110,-20){(a)}
%\put(125,-20){(b)}
\end{center}
\vspace{-2mm}
\caption{Behaviors of (a) the dilaton field $\phi$, (b) the mass function
$\tm_g/\tilde{M}_0$ and (c) the lapse function $\delta$ of the black hole solutions
with $k=-1$ in $D=10$. The horizon radii  and the masses are
$\tr_H =2.4$, $\tilde{M}_0=-1.8787\times 10^{1}$ (solid line),
$\tr_H =2.6$, $\tilde{M}_0=3.0737\times 10^{1}$ (dashed line),
$\tr_H=3.0$, $\tilde{M}_0=5.5498\times 10^{2}$ (dotted line), and
$\tr_H=4.0$, $\tilde{M}_0=1.7587\times 10^{4}$
(dot-dashed line).
}
\label{d10-_config}
\end{figure}

We show the configurations of the fields in the black hole solutions
in Fig.~\ref{d4-_config} -- \ref{d10-_config}.
For large black holes ($\tr_H>\tilde{\ell}_{\rm AdS}$), the dilaton field increases
monotonically to its asymptotic value $\phi_0$. The mass function in $D=4$ decreases
while those in higher dimensions $D\geq 5$ increase monotonically.
For $\tr_H=\tilde{\ell}_{\rm AdS}$, the functions $\phi$, $m_g$ and $\delta$
are constant.
For small black holes ($\tr_H<\tilde{\ell}_{\rm AdS}$), the dilaton
field decreases and becomes steep around the event horizon
as $\tr_H$ approaches its critical radius.
The mass function in $D=4$ behaves monotonically while there is a range
where the mass function changes its behavior in  $D\geq 5$.

% figures -------------------------------------------------------------------
% figures ---------------------------------------------------------------------

% figures -------------------------------------------------------------------
% figures ---------------------------------------------------------------------

% figures -------------------------------------------------------------------
% figures ---------------------------------------------------------------------

% figures -------------------------------------------------------------------
% figures ---------------------------------------------------------------------

%%%%%%%%%%%%%%%%%%%%%%%%%%%%%%%%%%%%%%%%%%%%%%%%%%%%%%%%%%%%%%%%
\section{Conclusions and Discussions}
\label{CD}
%%%%%%%%%%%%%%%%%%%%%%%%%%%%%%%%%%%%%%%%%%%%%%%%%%%%%%%%%%%%%%%%

We have studied the black hole solutions in dilatonic Einstein-GB theory with
the negative cosmological constant. The cosmological constant introduces
the Liouville type of potential for the dilaton field with a certain coupling.
We have studied the spherically symmetric ($k=1$) and hyperbolically symmetric
($k=-1$)  spacetimes. The basic equations
have some symmetries which are used to generate the black hole solutions
with different horizon radii and the cosmological constants.

The black hole solution should have a regular event horizon
and be singularity free in the outer region.
By the asymptotic expansion at infinity, where the spacetime approaches
AdS spacetime, the power decaying rate of the fields are estimated.
We have imposed the condition that the ``mass"
of the dilaton field satisfies the BF bound, which guarantees the stability of
the vacuum solution. By this condition, the values of the dilaton coupling
constant and the parameter of the Liouville potential are constrained.
For a typical choice of the parameters and boundary conditions, we were able
to construct AdS black hole solutions in various dimensions numerically.

We have chosen $\gamma=1/2$ and
$\lambda=1/3$ for the actual numerical analysis. The black hole solutions
are constructed in $D=4, 5, 6$ and $10$. We have checked that the dilaton field
climbs up its potential slope and takes constant values at infinity.
The field equations have exact solutions, i.e.,
regular AdS solution for $k=1$ and a massless black hole solution for $k=-1$.
The nontrivial solutions are obtained numerically in $D=4$ -- 6
and 10 dimensional spacetimes.

For spherically symmetric solution ($k=1$), there is the critical horizon radius
below which no solution exists in $D=4$ -- 6. In $D=4$, The field
functions diverge at the horizon for solution with the critical horizon radius.
In  $D=5$ and 6, the event horizon of the critical solution is regular but
the derivatives of fields diverge at some radius outside the event horizon.
In $D=10$, however, there is not such lower bound of the horizon radius but
the solution continues to exist to zero horizon size.
In the large black hole limit ($\tr_H\to \infty$), the solution approaches the planar
symmetric one ($k=0$), and the mass is proportional to $\tr_H^{D-1}$.

For hyperbolically symmetric solution ($k=-1$), the solutions are classified
by their horizon radius as $\tr_H>\tilde{\ell}_{\rm AdS}$ or
$\tr_H<\tilde{\ell}_{\rm AdS}$.
As the $\tr_H$ becomes large,  $\phi_H$ decreases, and approaches $\phi_H^{(k=0)}$
in the $\tr_H\to \infty$ limit.
The gravitational mass $\tilde{M}_0$ is monotonic with respect to $\tr_H$, and
is proportional to $\tr_H^{D-1}$.
The solution with $\tr_H=\tilde{\ell}_{\rm AdS}$ is given by Eq.~\p{AdSBH}, which is the
zero mass black hole.
For the solution with $\tr_H<\tilde{\ell}_{\rm AdS}$, $\phi_H$ decreases
as the horizon radius becomes small. There is the critical horizon radius
in all dimensions, and the fields of the critical solution diverge at the horizon.
The mass of these solution are negative.

\begin{figure}[t]
\begin{center}
\includegraphics[width=15cm]{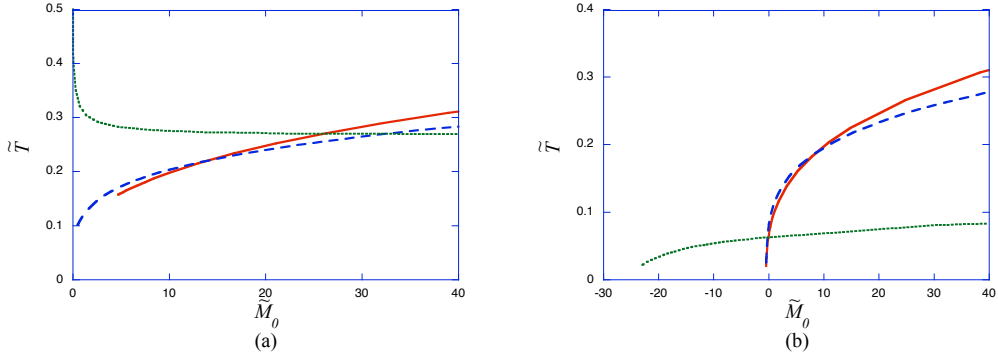}
%\put(-110,-20){(a)}
%\put(125,-20){(b)}
\end{center}
\vspace{-2mm}
\caption{
Temperature of the black hole solution with
(a) $k=1$ and (b)  $k=-1$ in
$D=4$ (solid line),
$D=5$ (dashed line),  and
$D=10$ (dotted line).
}
\label{temperature}
\end{figure}

There are some remaining issues left for future works.
One of them is the thermodynamics of our black holes.
The Hawking temperature is given by the periodicity of the Euclidean time
on the horizon as
\bea %----------------
\label{temp}
\tilde{T}_H ~&& =\frac{e^{-\d_H}}{4\pi}B_H'
\nonumber \\
&& =\frac{e^{-\d_H}}{4\pi h_H}
\biggl[\frac{(D-3)k}{\tr_H} +\frac{(D-3)_5k^2}{\tr_H^2} e^{-\c\phi_H}
- \frac{ \tr_H\tilde\Lambda e^{\la\phi_H}}{D-2}\biggr].
\ena %----------------
In the case of GB gravity, the entropy is not obtained by a quarter of the area
of the event horizon.
Along the definition of entropy in Ref.~\citen{Wald}, which originates from
the Noether charge associated with the diffeomorphism invariance of
the system, we obtain
\bea %----------------
\tilde{S} = \frac{\tr_H^{D-2}\Sigma_{k}}{4}
\left[1+2(D-2)_3 \frac{ke^{-\c\phi_H}}{\tr_H^2} \right]-\tilde{S}_{min},
\ena %----------------
where $S_{min}$ is added to make the entropy non-negative\cite{Clunan}.
The temperatures of the black holes are plotted in Fig.~\ref{temperature}.
In $D=10$, the temperature of the black hole with $k=1$ diverges
in the zero horizon radius limit while the other solutions have finite temperature
even for the solutions with the critical horizon radius.
For the solutions with $k=1$ in $D=4$ and the ones with $k=-1$, the critical
horizon radius is determined by the reality of $\phi'$ at horizon.
At the horizon of the critical solution, $\phi''$ diverges but $B'$ and $\delta$
are finite as discussed in \S~\ref{spherical} and \S~\ref{hyperbolic}.
For the solutions with $k=1$ in $D=5$ and 6, the critical solutions
are determined by the divergence of the second derivative of the
dilaton field outside the horizon. This means that the horizon is regular.
Hence for these solutions the temperature is finite.

% figures -------------------------------------------------------------------
% figures ---------------------------------------------------------------------

Since the temperature of the black hole is non-zero
for all mass range, we expect that our black holes
will not stop evaporating process via Hawking radiation.
Then the following fates are considered:
(i) The black hole becomes small and the horizon radius reaches
the critical radius. Then the naked singularity appears.
(ii) The solution is deformed to less symmetric (non-spherically/hyperbolically symmetric)
one.
(iii) Other higher curvature correction terms become important and
lead to different fates from the present analysis.
Since the curvatures diverge for the critical solution, the case (iii) seems
plausible.
However we have to study the evaporation process
more carefully, and let us examine the process in the present model.

Recall that the temperature of the extreme black hole with
$\gamma >\sqrt{2}$ in the  Einstein-Maxwell-Dilaton system is infinite\cite{GM,GHS}.
Although the naive expectation is that this leads to an infinite emission rate,
Holzhey and Wilczek showed that the potential, through which created particles
travel  away to infinity, grows infinitely high in the extreme
limit, and then it is expected that the emission rate could be suppressed
to a finite value\cite{HHW} although the detailed analysis leads to the
different conclusion.~\cite{Koga}
In  our case, if the potential barrier becomes infinitely large, the emission rate
might be suppressed to zero even though the temperature of the black
hole remains finite. Then evaporation may stop and the black hole cannot reach
the critical solution.
To study this possibility, we have to calculate the potential barrier for
the field of evaporating particles in the background of our new solutions.

Here we examine a free neutral massless scaler field $\Phi$, which obeys
the Klein-Gordon equation:
%----------------------------
\begin{equation}
\squaret \:\Phi =0.
\label{4-1}
\end{equation}%---------
We expand the scalar field in harmonics and study the mode
%----------------------------
\begin{equation}
\Phi = \frac{\chi(\tr)}{\tr^{(D-2)/2}} Y_{lm}(\psi) e^{-i\omega t}.
\label{4-2}
\end{equation}%---------
Eq. (\ref{4-1}) becomes separable, and the radial equation reduces to
%----------------------------
\begin{equation}
\left[ \frac{d^2}{d\tr^{\ast 2}} + \omega^2 - V^2 (\tr) \right] \chi(\tr^
{\ast}) =0,
\label{4-3}
\end{equation}%---------
where $\tr^{\ast}$ is the tortoise coordinate defined as
%----------------------------
\begin{equation}
\frac{d}{d\tr^{\ast}} = Be^{-\delta} \frac{d}{d\tr}
\label{4-4}
\end{equation}%---------
and $V^2 (r)$ is the potential;
\begin{equation}
V^2 (\tr) = \frac{(D-2)Be^{-2\delta}}{2\tr^2}
      \left[ l(l+D-3) +\tr B' -\tr\delta'B+\frac{D-4}{2}
      \right].
\label{4-5}
\end{equation}

\begin{figure}[t]
\begin{center}
\includegraphics[width=9cm]{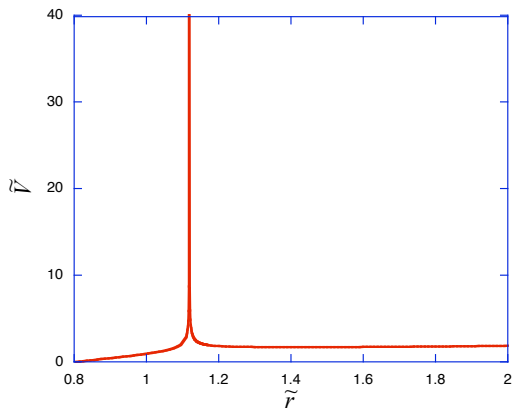}
\end{center}
\vspace{-2mm}
\caption{
Potential function of the free massless scalar field around the near critical black
hole solution with $k=1$ and $\tr_H=0.805$ in $D=5$.
There is a sharp barrier outside the event horizon.
}
\label{potential}
\end{figure}

For the critical solutions with $k=1$ in $D=4$ and the ones with $k=-1$,
the curvature singularity appears at the horizon, where $\delta'$ diverges
but $B=0$. Hence the potential is finite there. As a result, the evaporating
process does not stop.
On the other hand, for the solutions with $k=1$ in $D=5$ and 6,
the potential of the near critical black hole is plotted in Fig.~\ref{potential}.
The potential forms sharp barrier and it may stop the evaporation and give
a regular remnant.

To study the evolution of the black hole, deeper consideration should be
necessary including the higher order corrections and potential barrier at
AdS infinity. This is left for future study.
A related issue is that the existence or non-existence of these critical solutions
seems to depend on the value of the parameter $\c$ and $\la$.~\cite{MOS}
This is expected because the dilaton significantly affects the solutions
if the dilaton coupling is stronger.
Stability and cosmological implications of these solutions should also be
studied.~\cite{PC}
The solution for the positive cosmological constant is also a subject of future
study, which will be reported elsewhere.
Finally we hope that our asymptotically AdS black hole solutions are
useful for examining properties of field theories via AdS/CFT correspondence.
% figures -------------------------------------------------------------------
% figures ---------------------------------------------------------------------

\section*{Acknowledgements}
We would like to thank Z. K. Guo for collaboration on the earlier papers,
and Kei-ichi Maeda for valuable discussions.
The work of N.O. was supported in part by the Grant-in-Aid for
Scientific Research Fund of the JSPS Nos. 20540283 and 06042,
and also by the Japan-U.K. Research Cooperative Program.

%%%%%%%%%%%%%%%%%%%%%%%%%%%%%%%%%%%%%%%%%%%%%%%%%

%%%%%%%%%%%%%%%%%%%%%%%%%%%%%%%%%%%%%%%%%%%%%%%%%%%%

\end{document}